\documentclass[10pt,journal,compsoc,peereview]{IEEEtran}
\usepackage[space,sort,nocompress]{cite}
\usepackage{amsmath,amssymb,amsfonts}
\usepackage{algorithmic}
\usepackage{graphicx}
\usepackage{textcomp}
\usepackage{xcolor}
\usepackage{booktabs,chemformula}
\usepackage{booktabs}       
\usepackage[switch]{lineno}
\usepackage{fancyhdr}
\usepackage[hyphens]{url}

\usepackage{caption}
\usepackage{subcaption}
\definecolor{Highlight}{rgb}{0.0, 0.0, 0.0}

\ifCLASSOPTIONcompsoc
  \usepackage[nocompress]{cite}
\else
  \usepackage{cite}
\fi
\ifCLASSINFOpdf
\else
\fi
\begin{document}

%
\title{Dissecting Tensor Cores via Microbenchmarks: Latency, Throughput and Numeric Behaviors }

\author{Wei Sun,
        Ang Li, 
        Tong Geng, 
         Sander Stuijk, 
         Henk Corporaal 
\IEEEcompsocitemizethanks{\IEEEcompsocthanksitem Wei Sun, Sander Stuijk and Henk Corporaal are with Electronic System Group, Eindhoven University of Technology, the Netherlands.
\protect\\
E-mail: \texttt{\{w.sun,s.stuijk,h.corporaal\}@tue.nl}
\IEEEcompsocthanksitem Ang Li and Tong Geng are with the Physical and Computational Sciences Directorate, Pacific Northwest National Laboratory, WA, USA. 
\protect\\
E-mail: \texttt{ \{ang.li,tong.geng\}@pnnl.gov} 
}
\thanks{Manuscript received May 26, 2022; revised October 21, 2022.}}

\IEEEtitleabstractindextext{%
\begin{abstract}

Tensor Cores have been an important unit to accelerate Fused Matrix Multiplication Accumulation (MMA) in all NVIDIA GPUs since Volta Architecture. To program Tensor Cores, users have to use either legacy wmma  APIs or current mma APIs. Legacy wmma APIs are more easy-to-use but can only exploit limited features and power of Tensor Cores. Specifically, wmma APIs support fewer operand shapes and can not leverage the new sparse matrix multiplication feature of the newest Ampere Tensor Cores. However, the performance of current programming interface has not been well explored.
Furthermore, the computation numeric behaviors of low-precision floating points (TF32, BF16, and FP16) supported by the newest Ampere Tensor Cores are also mysterious. In this paper, we explore the throughput and latency of current programming APIs. We also intuitively study the numeric behaviors of Tensor Cores MMA and profile the intermediate operations including multiplication, addition of inner product, and accumulation. \textcolor{Highlight}{All codes used in this work can be found in \url{https://github.com/sunlex0717/DissectingTensorCores}}.

\end{abstract}

\begin{IEEEkeywords}
GPU, Tensor Cores, Numeric profiling, Ampere, Turing, Microbenchmark
\end{IEEEkeywords}}

\maketitle

\IEEEdisplaynontitleabstractindextext

\IEEEpeerreviewmaketitle

\IEEEraisesectionheading{\section{Introduction}\label{sec:introduction}}


\IEEEPARstart{G}{eneral} Matrix Multiplication (GEMM) is a fundamental computation pattern in modern HPC applications and especially deep learning applications. To meet the increasing demands of high throughput GEMM, many commercial hardware accelerators have been designed such as NVIDIA Tensor Cores~\cite{VoltaWhite}, Google TPUs~\cite{TPU}, Intel Nervana~\cite{hickmann2020intel} and Xilinx Versal AI Engines~\cite{gaide2019xilinx}. 
NVIDIA introduced the Tensor Cores Unit first in its Volta Architecture~\cite{VoltaWhite} and integrated this special computation unit in the GPGPU architecture. Tensor Cores can provide significant speed-up compared to traditional CUDA cores and more numeric precision choices to satisfy different demands~\cite{SC18Harnessing,micikevicius2017mixedTraining,li2020acceleratingBNN,feng2021apnnSC21}.

NVIDIA so far has released three generations of Tensor Cores -  Volta~\cite{VoltaWhite}, Turing~\cite{TuringWhite}, and Ampere~\cite{AmpereWhite}. Now Tensor Cores are becoming the standard components of all NVIDIA GPU products including high-end server GPUs (e.g. V100 and A100), Gaming GPUs (e.g. RTX20xx and RTX30xx) as well as Embedded GPUs (e.g. Jetson Xavier and Orin). 
However, the programming model and behaviors of the Tensor Cores are significantly different from previously well-explored \textbf{CUDA cores}. Furthermore, there are also noticeable differences between different Tensor Cores generations. For instance, Ampere Tensor Cores support fine-grained N: M sparse acceleration for Sparse Matrix Multiplication, which can only be programmed via the new mma instructions instead of previously studied legacy wmma instructions~\cite{markidis2018NVIDIA,raihan2019modeling}.
There are also some incompatible instructions between Ampere and prior generations. For example, Ampere Tensor Cores do not support the {HMMA.884} Assembly instruction; the corresponding mma.m8n8k4 PTX instruction is essentially compiled into a set of Floating point unit (FPU) Assembly instructions that are 10x slower than the expected Tensor Cores performance. In contrast, previous observations of the Volta Tensor Cores~\cite{raihan2019modeling} suggest that this HMMA.884 is the fundamental machine code (SASS), and all legacy wmma.mma instructions will be compiled into a set of HMMA.884 SASS codes. Therefore, it is vital to understand the programming model and behavior of the Tensor Cores for achieving the best possible performance on different GPU architectures.


Despite the previous efforts on microbenchmarking Tensor Cores~\cite{markidis2018NVIDIA,raihan2019modeling,martineau2018benchmarkingV100,yan2020demystifying,fasi2021numerical,jia2018dissectingvolta,jia2019dissectingturing} from performance and numeric perspectives. There is no existing work that evaluates the current programming APIs with instruction-level microbenchmarking and numeric studies. Table~\ref{table:summary-tc} in Section~\ref{sec:related-work} compares our work with previous Tensor Cores studies. Current programming APIs have significant advantages over legacy wmma APIs such as more flexible operand shapes, better performance, and new sparse acceleration on Ampere Tensor Cores. Therefore, a comprehensive and up-to-date study based on the new programming interface is necessary to complement the findings of existing literature. Our research fulfills this gap by making the following contributions:


\begin{itemize}
    \item A new set of microbenchmarks to explore the three core PTX instruction sets closely related to Tensor Cores --- ldmatrix, mma, and mma.sp. 
    
    \item Benchmarking the Tensor Cores instruction latency and throughput. We provide in-depth analysis and provide  programming guidelines which are needed to implement custom applications on Tensor Cores that are not well supported by vendor libraries.  
    
    
    \item Profiling the numeric behavior of low-precision floating data types supported in newest Ampere Tensor Cores - TF32, BF16, and FP16. 
    
\end{itemize}

Our work provides comprehensive and up-to-date information on the Tensor Cores. Our microbenchmarks cover the programming interface, latency, and throughput of Tensor Cores instructions as well as numeric behaviors of low-precision floating point operations. To the best of our knowledge, this paper is the first systematic study on recent Tensor Cores generations (Turing and Ampere) using the new programming interface. 


The rest of the paper is organized as follows: Section~\ref{sec:Background} introduces the background knowledge of NVIDIA GPUs and Tensor Cores. Section~\ref{sec:related-work} summarizes the related work and highlights the differences between our work. Section~\ref{sect:methodology} introduces the microbenchmark methodology used in this paper. Section~\ref{sect:DesenTC},~\ref{sect:SparseTC} and~\ref{sect:DataMovement} microbenchmark the performance of mma, mma.sp and ldmatrix instructions respectively. Section~\ref{sec:numeric} presents our numeric studies. Section~\ref{sec:conclusion} concludes our findings. 

 
\section{Background of Tensor Cores GPUs}\label{sec:Background}

We first introduce the general architecture of modern Tensor Cores GPU in Section~\ref{subsec:GPU-arch-background}, and then we compare the three Tensor Cores generations and highlight their differences in Section~\ref{subsec:TensorCoreCompare}.

\subsection{Background of Modern GPU Architecture}\label{subsec:GPU-arch-background}

Tensor Cores is a domain-specific computation unit integrated into NVIDIA GPUs for accelerating matrix multiplication (MM), which performs D = A$\times$B + C, where A and B are input matrices with the shape of m$\times$k and k$\times$n, respectively; C is accumulator and D is the result. 

\begin{figure}[htp]
    \centering
    \includegraphics[width=8cm]{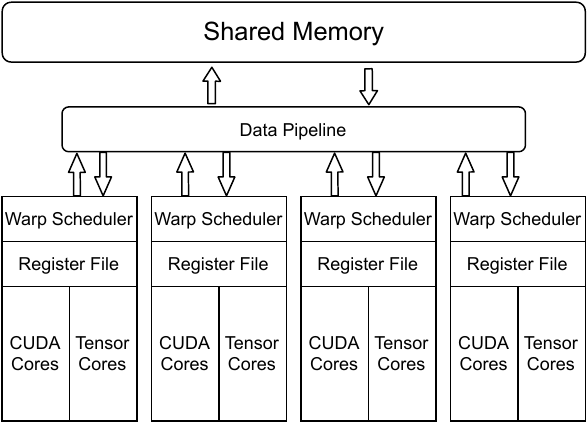}
    \caption{ Simplified architecture overview of Tensor Cores GPU SM~\cite{AmpereWhite,TuringWhite,VoltaWhite}. Each SM consists of four sub-warps. Each sub-core has its warp scheduler, register file, CUDA cores, and Tensor Cores.}
    \label{fig:uArch}
\end{figure}

Modern GPU architecture consists of a certain amount of Streaming Multiprocessor (SM) which works as the fundamental processing unit of the GPU. These SMs are further connected to the GPU device memory/global memory which is shared by all SMs. Each SM has its internal caches, Shared Memory, Register File, CUDA cores, Tensor Cores, Load Store units, and other special units. Figure~\ref{fig:uArch} illustrates the abstract SM architecture of Tensor Cores integrated GPUs. Each SM consists of four warp schedulers or four sub-cores to issue four warp instructions simultaneously. Tensor Cores use the same memory hierarchy as CUDA cores. Both Tensor Cores and CUDA cores can fetch data through direct addressing ($ld$ instruction) from the Shared Memory or global memory via the \textbf{per-thread}\footnote{Per-thread means the instruction is executed by a single thread independently; both behavior and result of the thread are deterministic.} scheme. Tensor Cores have two special load instructions -- recent $ldmatrix$ and legacy $wmma.load$ via the \textbf{per-warp}\footnote{Per-warp means the instruction is executed by the 32 threads within the same warp cooperatively; neither behavior nor the result of individual thread is deterministic.} scheme.

In this paper, we exclude the microbenchmark experiments and discussions on global memory access because 1) Using Shared Memory as the buffer for global memory to increase data reuse has been a standard optimization technique for accelerating GEMM-like applications on GPUs.
2) The novel asynchronous global memory copy introduced in Ampere Architecture facilitates the software data pipeline by using Shared Memory as the staging storage to overlap the computation with the data transfer from global memory to Shared Memory~\cite{AmpereWhite}. 3) The ldmatrix PTX instruction can only load data from Shared Memory. Therefore, without losing generality, we assume the input data has already been fetched into the Shared Memory from global memory since our goal is to microbenchmark the Tensor Cores. \textcolor{Highlight}{We present ablation experiments in Appendix \ref{appendix:sharedmemUsage} to further demonstrate the advantages of using shared memory and asynchronous copy.}


\subsection{The Evolution of Tensor Cores}\label{subsec:TensorCoreCompare}

Up to date, there are three generations of Tensor Cores in the market --- Volta, Turing, and Ampere. NVIDIA recently announced the fourth generation of Tensor Cores GPU Architecture known as Hopper~\cite{HopperWhite}. However, Hopper GPUs are not publicly released yet.

{Table~\ref{table:TensorCoresCompare}} compares the three released Tensor Cores generations and preliminary features of the fourth-generation Hopper Tensor Cores. Volta Tensor Cores are the first generation and support only FP16 as the input data type. Turing Tensor Cores are the enhanced version of Volta Tensor Cores and support three extra data types -- INT8, INT4, and Binary~\cite{li2020acceleratingBNN}. The third-generation Ampere Tensor Cores introduce acceleration for sparse matrix multiplication with fine-grained structured sparsity and a new machine learning data type called bfloat16 (BF16). Furthermore, Ampere Architecture redesigns the micro-architecture of Tensor Cores. Unlike Volta and Turing Architecture which have eight Tensor Cores per SM and each Tensor Core performs a 4$\times$4$\times$4 MM (i.e. m = n = k = 4), there are only four Tensor Cores per SM and each Tensor Core performs an 8$\times$4$\times$8 MM. The fourth-generation Hopper will bring a new 8-bit floating-point numeric precision (FP8) for accelerating certain machine learning applications that exhibit higher tolerance for training based on low-precision data types. Furthermore, it seems INT4 and Binary are no longer discussed in the Hopper Whitepaper~\cite{HopperWhite}. Section~\ref{sec:numeric} will give more discussions on the numeric behaviors of low-precision floating-point data types of Tensor Cores (i.e. TF32,BF16 and FP16).


\begin{table*}[ht]
\caption{Comparisons of the properties of different generations of Tensor Cores. Note that Hopper GPU is not available on the market at this moment, the features are preliminary and could be adjusted according to the vendor's documentation~\cite{HopperWhite}. \textbf{NA} means not being documented by the vendor yet.}
\label{table:TensorCoresCompare}
\centering

\begin{tabular}{|c|c|c|c|c|c|c|}
\hline
Architecture & Representative products   & Numeric types                                                                   & TCs/SM & m$\times$n$\times$k & Sparse Acceleration        & Programmability                                                      \\ \hline
Volta        & V100, Jetson Xavier       & FP16                                                                            & 8      & 4$\times$4$\times$4 & No                         & wmma                                                                 \\ \hline
Turing       & T4,RTX20x                 & FP16, INT8, INT4, Binary                                                          & 8      & 4$\times$4$\times$4 & No                         & \begin{tabular}[c]{@{}l@{}}wmma, ldmatrix,\\ mma\end{tabular}        \\ \hline
Ampere       & A100, RTX30x, Jetson Orin & \begin{tabular}[c]{@{}l@{}}FP16, BF16, TF32,\\ FP64, INT8, INT4, Binary\end{tabular} & 4      & 8$\times$4$\times$8 & fine-grained 50\% sparsity & \begin{tabular}[c]{@{}l@{}}wmma, ldmatrix,\\ mma, mma.sp\end{tabular} \\ \hline
Hopper       & H100                      & \begin{tabular}[c]{@{}l@{}}FP16, BF16, TF32,\\ FP64, FP8, INT8\end{tabular}         & 4      & NA    & fine-grained 50\% sparsity & \begin{tabular}[c]{@{}l@{}}wmma, ldmatrix,\\ mma, mma.sp\end{tabular} \\ \hline
\end{tabular}
\end{table*}


\begin{figure}[htp]
    \centering
    \includegraphics[width=8cm]{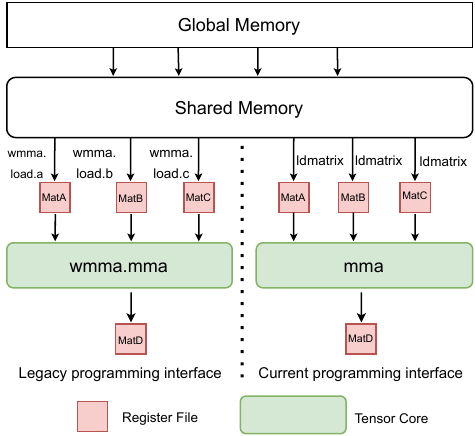}
    \caption{Legacy and current programming interface for Tensor Cores.}
    \label{fig:TCoverview}
\end{figure}

The programming of Tensor Cores has also evolved over the generations. Figure~\ref{fig:TCoverview} illustrates the legacy and current programming interface for Tensor Cores. The first step is to load data from Shared Memory to the Register File and then use Tensor Cores to accelerate the matrix computation. Legacy wmma interface was introduced with first-generation Volta Tensor Cores which have been extensively studied by previous work~\cite{markidis2018NVIDIA,raihan2019modeling,martineau2018benchmarkingV100,yan2020demystifying}. In general, the legacy wmma programming interface can satisfy basic usage of Tensor Cores and requires less programming efforts, because wmma.load can help manage the special input operand storage layout in the Register File for Tensor Cores. However, wmma instructions can only leverage limited features of the Tensor Cores~\cite{liu2021optimizing}. Specifically, it can not use sparse acceleration and has fewer choices of the matrix operand shape. Furthermore, wmma.load instructions have more strict requirements for the data layout in Shared Memory which will be introduced in Section~\ref{sect:DataMovement}. By contrast, the new programming interface provides access to all features of the Tensor Cores (i.e. more matrix operand shapes and novel sparse acceleration). The vendor's library CUTLASS~\cite{CUTLASS} also chose the new programming interface as the underlying implementation for the best possible performance. Therefore, users should use the new programming interface when seeking for best possible performance or exploiting the new sparse acceleration feature.


\begin{figure}[htp]
    \centering
    \includegraphics[width=8cm]{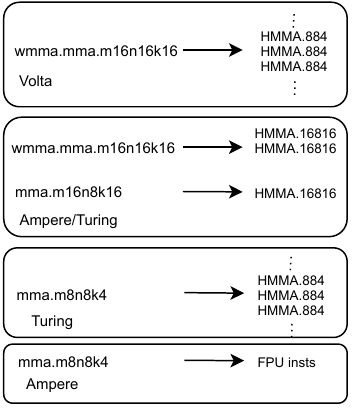}
    \caption{The compilation from legacy and current PTX instructions to SASS assembly codes~\cite{jia2018dissectingvolta,jia2019dissectingturing,raihan2019modeling,yan2020demystifying}.}
    \label{fig:compile}
\end{figure}

Figure~\ref{fig:compile} uses an example to illustrate the differences between legacy wmma.mma and current mma PTX instructions. Previous studies~\cite{raihan2019modeling,yan2020demystifying,jia2018dissectingvolta} reveal that on Volta Tensor Cores, every wmma.mma instruction will be compiled to a set of HMMA.884 SASS instruction. By contrast, when using a mma instruction on Ampere or Turing Tensor Cores, it will be compiled to a single HMMA SASS instruction. On the other hand, when attempting to use the legacy wmma.mma on Turing/Ampere Tensor Cores, it will be compiled into several new HMMA instructions~\cite{jia2019dissectingturing}. In the example of Fig.~\ref{fig:compile}, one legacy wmma.mma.m16n16k16 is complied into two HMMA.16816 (corresponding to new mma.m16n8k16) instructions. Note that there is one special instruction mma.m8n8k4. On Turing Tensor Cores, it will be compiled into a couple of HMMA.884 instructions which are similar to Volta Tensor Cores. However, on Ampere Tensor Cores, it will be compiled into a set of FPU instructions which will eventually run on the CUDA cores and lead to inferior performance than expected for the Tensor Cores.  

\section{Related Work}\label{sec:related-work}


\begin{table*}[htp]
\caption{Comparisons with previous Tensor Cores microbenchmark literature. Note that~\cite{jia2019dissectingturing} and~\cite{jia2018dissectingvolta} are general microbenchmark studies and cover Tensor Cores as one Section. Others~\cite{markidis2018NVIDIA,raihan2019modeling,martineau2018benchmarkingV100,yan2020demystifying,fasi2021numerical} are focusing on Tensor Cores. \cite{fasi2021numerical} only studies the numeric behaviors and performance is not evaluated. Our work is the first comprehensive study that evaluates current mma programming APIs and sparse Tensor Cores with instruction-level microbenchmarking and numeric studies.}
\label{table:summary-tc}
  \centering
          \small
        \setlength\tabcolsep{2pt}
  \begin{tabular}{cccccccc}
    \toprule
    Literature & GPU Arch   & Dense FMA & Sparse FMA & Data movement & Performance evaluations   & Numeric studies   & Assembly studies \\ 
    \midrule
    \cite{markidis2018NVIDIA} & Volta   & wmma.mma & --  & wmma.load  & \begin{tabular}[c]{@{}l@{}}Vendor libraries \\ benchmark\end{tabular}  & FP16 &  -- \\ 
    \midrule
    \cite{martineau2018benchmarkingV100} &  Volta                                                           & wmma.mma & --  & wmma.load  & \begin{tabular}[c]{@{}l@{}}Vendor libraries \\ benchmark\end{tabular}& --    & --   \\ 
    \midrule
    \cite{jia2018dissectingvolta} & Volta    & wmma.mma & --      & wmma.load  &
    \begin{tabular}[c]{@{}l@{}}Vendor libraries \\ benchmark\end{tabular} & --  & $\checkmark$  \\
    \midrule
    \cite{jia2019dissectingturing} & Turing & wmma.mma &--     & wmma.load  & \begin{tabular}[c]{@{}l@{}}Vendor libraries \\ benchmark\end{tabular}     & --   & $\checkmark$   \\
    \midrule
    \cite{raihan2019modeling} & \begin{tabular}[c]{@{}l@{}}Turing \\ Volta\end{tabular} & wmma.mma & --  & wmma.load  & \begin{tabular}[c]{@{}l@{}}Instruction-level\\ microbenchmark\end{tabular} & -- & $\checkmark$ \\ 
    \midrule
    \cite{yan2020demystifying} & \begin{tabular}[c]{@{}l@{}}Turing \\ Volta\end{tabular}  & wmma.mma & --  & wmma.load  & \begin{tabular}[c]{@{}l@{}}Instruction-level\\ microbenchmark\end{tabular} & -- & $\checkmark$  \\ 
    \midrule
    \cite{fasi2021numerical} &\begin{tabular}[c]{@{}l@{}}Ampere\\ Turing\\ Volta\end{tabular} & wmma.mma & -- & wmma.load  & --  & \begin{tabular}[c]{@{}l@{}}TF32,BF16,\\ FP16\end{tabular} & --   \\ 
    \midrule
    \midrule
    Ours & \begin{tabular}[c]{@{}l@{}}Ampere \\ Turing\end{tabular}  & mma   & mma.sp  & ldmatrix   & \begin{tabular}[c]{@{}l@{}}Instruction-level\\ microbenchmark\end{tabular} &\begin{tabular}[c]{@{}l@{}}TF32,BF16,\\ FP16\end{tabular} & $\checkmark$  \\
    \bottomrule
  \end{tabular}
\end{table*}

GPU evaluations and microbenchmarks~\cite{wong2010demystifying,DissectMemory,jia2018dissectingvolta,jia2019dissectingturing,yan2020demystifying,markidis2018NVIDIA,li2019evaluating,raihan2019modeling,fasi2021numerical,martineau2018benchmarkingV100} are important technique to explore the unknown characteristics of GPU architecture like memory hierarchy, throughput, latency and numeric behaviors.

Table~\ref{table:summary-tc} summaries the existing Tensor Cores microbenchmark studies and compares them with our work.~\cite{markidis2018NVIDIA} studies the Volta Tensor Cores with legacy wmma programming interface and benchmarks the vendor libraries~\cite{cuBLAS,CUTLASS}. The authors also profile the numeric loss when using half (FP16) precision on Volta Tensor Cores. ~\cite{martineau2018benchmarkingV100} also benchmarks the vendor libraries and further studies the software optimization techniques when using wmma APIs to program Volta Tensor Cores. However, the experiments and evaluations of these two work~\cite {markidis2018NVIDIA,martineau2018benchmarkingV100} are only based on old Volta Tensor Cores. Furthermore, they do not provide instruction-level microbenchmarks (i.e. instruction latency and throughput) or discussions (e.g. Assembly codes studies).

~\cite{jia2018dissectingvolta} and~\cite{jia2019dissectingturing} explore the characteristics of Volta and Turing GPUs respectively. They cover the vendor libraries benchmarking and Assembly code studies of legacy wmma APIs for Tensor Cores. However, these two works are generic GPU studies. The discussions on Tensor Cores are limited; the numeric behaviors and instruction-level performance are not treated. 

~\cite{yan2020demystifying} demystifies Volta and Turing Tensor Cores through Assembly code benchmarking and further optimizes the half-precision matrix multiplication on Tensor Cores. The authors focus on exploring the Assembly level optimizations for matrix multiplication performance on Tensor Cores and achieve better performance than the vendor's library released at that moment. However, programming Tensor Cores via Assembly codes (SASS) is miserable since SASS codes are not officially documented by NVIDIA and there is not an official Assembler. Although Assembly codes may expose more optimization opportunities for certain applications~\cite{lavin2016fast}, users have to rely on third-party assemblers~\cite{yan2020winogradTuringass,zhang2017KepAs} developed through reverse engineering, which may be not stable and can be error-prone. In this work, we focus our studies on the PTX level, since PTX instructions are fully documented and supported by NVIDIA. 

~\cite{raihan2019modeling} studies the legacy wmma instructions on Turing and Volta Tensor Cores and proposes a Tensor Core model in GPGPU-SIM~\cite{bakhoda2009GPUSIM}, based on their findings that HMMA.884 is the fundamental execution pattern of the Tensor Cores. However, since there are significant differences between the newest Ampere and older Turing/Volta Tensor Cores as discussed in Section~\ref{subsec:TensorCoreCompare}, the behaviors and performance of Ampere Tensor Cores may not be simulated accurately by the old model.

~\cite{fasi2021numerical} focuses on the numeric behaviors of Tensor Cores and explores the underlying rounding modes and subnormal behaviors of TF32, BF16, and FP16 data types. However, the performance (i.e. latency and throughput) of Tensor Cores is not evaluated.

Compared with the aforementioned Tensor Cores studies, we make the following different contributions:

\begin{itemize}
\item We evaluate the current mma APIs (ldmatrix, mma, and mma.sp) instead of legacy wmma APIs. Current mma APIs provide more operand shapes and new sparse matrix multiplication acceleration, but have not been well studied by existing literature~\cite{markidis2018NVIDIA,jia2018dissectingvolta,jia2019dissectingturing,yan2020demystifying,fasi2021numerical,martineau2018benchmarkingV100,raihan2019modeling}. Our work fulfills this gap. 
\item We microbenchmark the instruction-level performance (i.e. throughput and latency). Compared to benchmarking vendor libraries~\cite{cuBLAS,CUTLASS}, our instruction-level studies can give insights and programming guidelines that the latency of different mma instructions with different operand shapes as well as how maximum performance can be achieved. This information is important to users who need to implement their applications, which are not well supported by libraries, on Tensor Cores~\cite{li2020acceleratingBNN,liu2021optimizing,feng2021apnnSC21,wang2021tc}.  
\item We intuitively study the numeric behaviors of three low-precision floating points (FP16, BF16, and TF32) w.r.t IEEE standard FP32 precision. We profile the numeric errors of three operations including multiplication, the addition of inner product, and accumulation. Our experiments complement findings in~\cite{fasi2021numerical} by comparing the error level and behaviors of the three data types, which can help users to decide which data types should be considered for their applications.

\end{itemize}


\section{Microbenchmark Methodology}\label{sect:methodology}


The goal of our microbenchmark is to offer the following performance information for each instruction:
\begin{itemize}
    
    \item[1)] \textit{Latency}: Latency is the duration (cycles) of starting to issue the instructions to the execution pipeline until the results become available for the next usage. Since GPU is a parallel architecture, there can be multiple instructions running simultaneously on each SM. When there is only one instruction per warp and one warp per SM, the measured latency is called \textit{completion/issue latency}, which reveals the length of the pipeline. 

    \item[2)] \textit{Throughput/bandwidth}: For data movement instruction, throughput is measured as bytes/clock cycle(clk)/SM. For computation instructions, throughput is FMA/clk/SM where FMA stands for fused multiplication accumulation. As the definition, d = a$\times$b+c counts as one FMA, and m$\times$n$\times$k matrix multiplication counts as m$\times$n$\times$k FMAs. Throughput/bandwidth can be measured by dividing the workload (i.e. bytes/SM or FMA/SM) by latency (cycles).

\end{itemize}

\begin{figure}[htp]
    \centering
    \includegraphics[width=8cm]{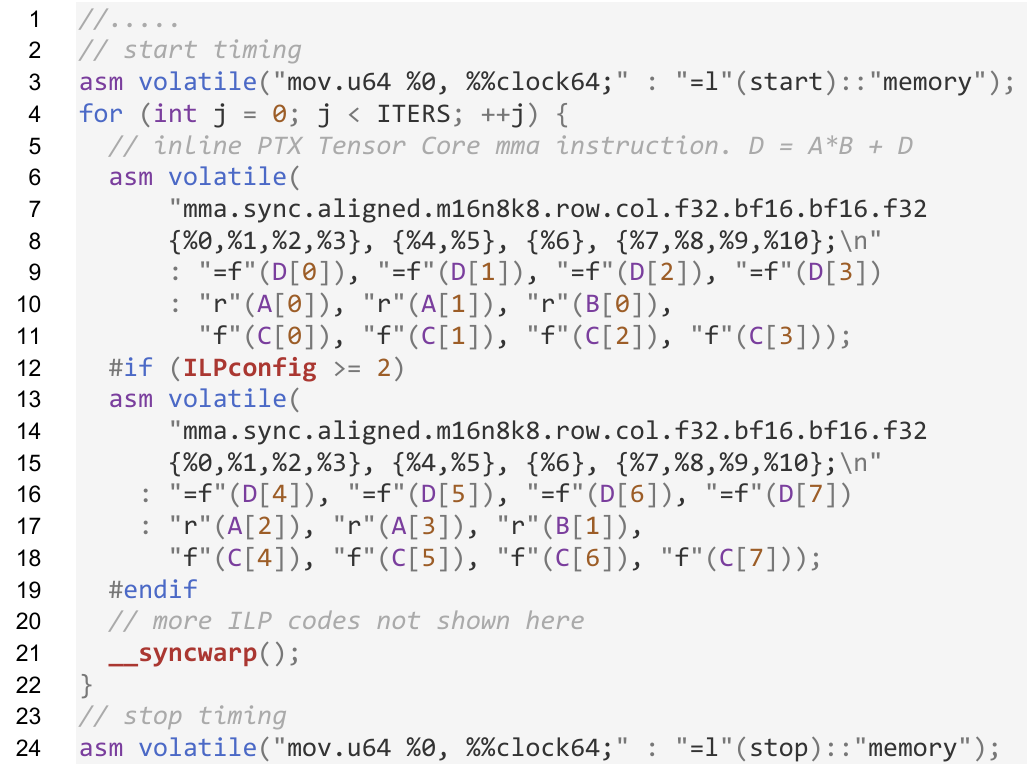}
    \caption{Code piece showing how to microbenchmark the performance of Tensor Cores. It contains a GPU timer (lines 3 and 24) and a for-loop to execute the body code mma instructions depending on the ILP configurations.}
    \label{fig:code-benchmark}
\end{figure}


Figure~\ref{fig:code-benchmark} shows the core code piece of how to measure the latency of the instruction with different Instruction Level Parallelism (ILP). Tensor Cores instructions (e.g. mma in the example) are executed in a per-warp scheme so we use warp-level synchronization at the end of each iteration to avoid optimizations (i.e. parallelism) across different iterations. The latency is then computed by taking the average of ITERS iterations. If ILP = 1 and we only launch one CUDA thread block with one warp (32 threads), the measured latency will be the instruction completion/issue latency. By increasing ILP and launching more warps per SM, we can measure the throughput and latency under different parallel configurations which can then give the architectural and programming insights of the Tensor Cores.

In summary, we will conduct our microbenchmark experiments for each instruction as follows:
\begin{itemize}
    \item[1)] Measure the completion/issue latency by choosing ILP = 1 and launching one warp per SM. 
    \item[2)] Measure the throughput and latency under different ILP and different warps per SM.
\end{itemize}



\section{Dense FMA}\label{sect:DesenTC}


As introduced in Section~\ref{subsec:TensorCoreCompare}, there are two approaches to program dense Tensor Cores - legacy wmma.mma and current mma instructions. We focus on the new mma instructions microbenchmarks since the legacy wmma.mma has been extensively researched by previous work~\cite{markidis2018NVIDIA,jia2018dissectingvolta,jia2019dissectingturing,yan2020demystifying,fasi2021numerical,martineau2018benchmarkingV100,raihan2019modeling}.



\begin{figure}[htp]
    \centering
    \includegraphics[width=8cm]{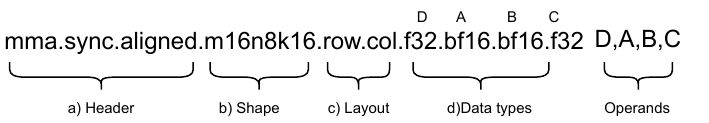}
    \caption{The mma instruction performs a D = A$\times$B + C MatMul. Segment m16n8k16 defines The shape of A(m$\times$k), B(k$\times$n), C(m$\times$n), and D(m$\times$n). The layout indicates the matrices are either row-major or col-major. The final instruction segment indicates the numeric types of D, A, B, and C respectively.}
    \label{fig:mma-inst}
\end{figure}

Figure~\ref{fig:mma-inst} shows a mma instruction example. According to the vendor's documentation~\cite{NVPTX}, for the BF16 data type, there are two choices of matrix shapes -- m16n8k8 and m16n8k16. Depending on the shapes, the BF16 mma PTX instructions will be compiled to HMMA.1688.FP32.BF16 or to HMMA.16816.FP32.BF16 SASS assembly code. Since the Tensor Cores support many different data types as shown in Table~\ref{table:TensorCoresCompare}, we present a detailed analysis for the BF16 data type on A100 Ampere Tensor Cores. The analysis for other data types is similar, so we provide the final results without losing generality.  




\subsection*{Experimental results}

\begin{figure}[htp]
     \centering
     \begin{subfigure}[*]{0.49\textwidth}
         \centering
         \includegraphics[width=\textwidth]{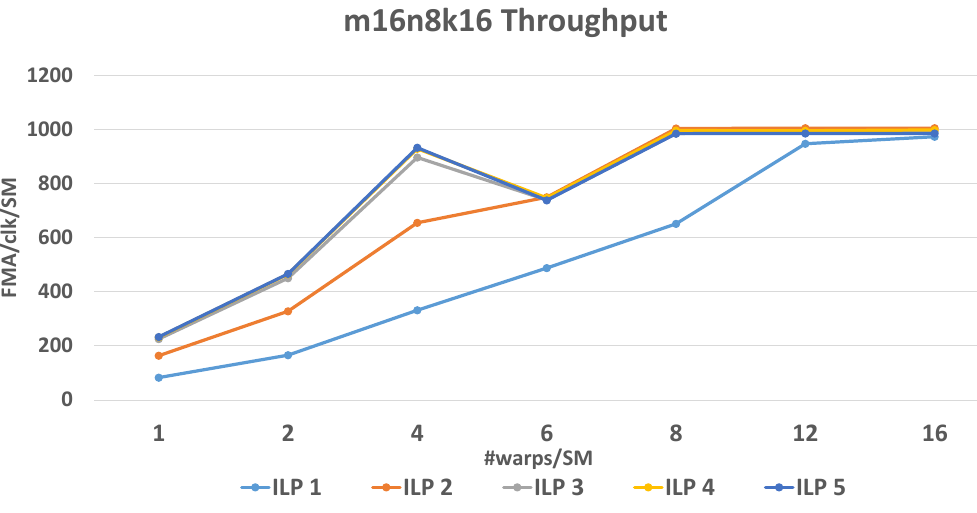}
         \label{fig:A100-m16n8k16-throughput}
     \end{subfigure}
    \begin{subfigure}[*]{0.49\textwidth}
         \centering
         \includegraphics[width=\textwidth]{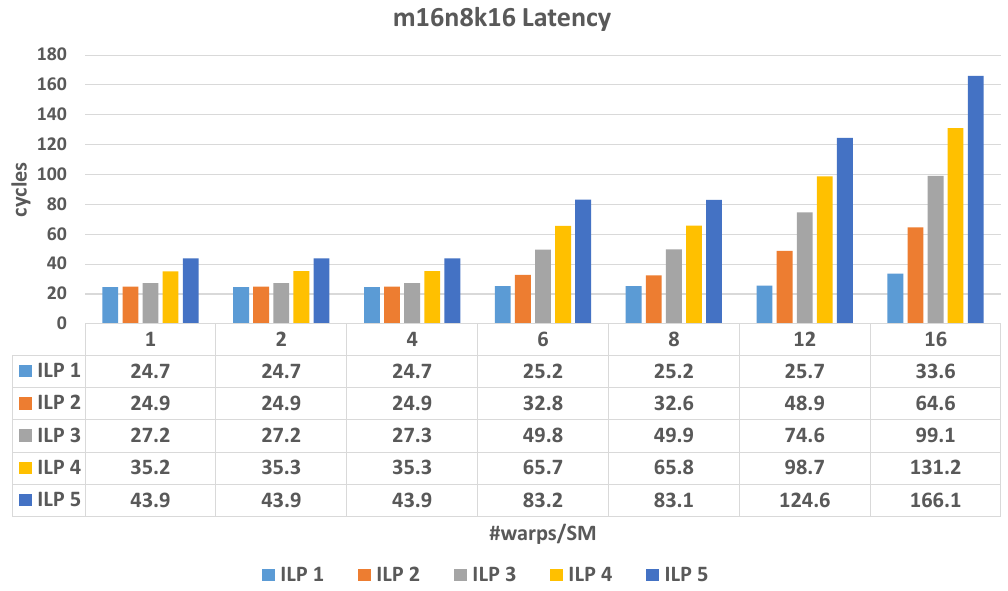}
         \label{fig:A100-m16n8k16-latency}
     \end{subfigure}
\caption{Throughput and latency of mma.m16n8k16 instruction under different settings on A100 Tensor Cores.}
\label{fig:A100-m16n8k16}
\end{figure}


Figure~\ref{fig:A100-m16n8k16} shows the measured throughput and throughput latency of mma.m16n8k16 instruction on A100 Ampere Tensor Cores GPU under different ILPs and \#warps/SM (shortened as \#warps). We start the experiments with ILP = 1 and \#warps = 1 and then increase the ILP and \#warps until convergence. Based on the results, we have findings as follows:

\begin{itemize}
   \item[1)] According to Section~\ref{sect:methodology}, the measured latency under ILP = 1 and \#warp = 1 is completion latency.
   The instruction completion latency is therefore around 25 (cycles). The measured peak throughput is 1000 (FMA/clk/SM), almost matching the number of 1024 as claimed by the vendor~\cite{AmpereWhite}.
    
    \item[2)] By analyzing the latency and throughput under \#warp = 1 and different ILPs, we observe that the peak throughput is 230 (FMA/clk/SM) 
    and converges at ILP = 3. Increasing ILP further will not lead to higher throughput but significantly more latency cycles.
    This is because the instructions of a specific warp \textbf{can not} use idle hardware resources located in other sub-cores even if they are within the same SM. One warp can only claim the resources of one sub-core.

    \item[3)] When \#warps $\leq$ 4 (1, 2, 4), the throughput scales with \#warps but the latency keeps unchanged. This observation confirms that the resources of the sub-core are isolated and each SM can issue four warps simultaneously. The maximum throughput can be achieved when \#warps = 8 and ILP $\geq$ 2. Note that \#warps = 4 with ILP $\geq$ 3 is also a convergence point, but it can not achieve the same performance as \#warps = 8, ILP =2 ( 900 v.s. 1000 ). This observation indicates that allocating more than 1 warp per sub-core helps overlap the intra-warp synchronization stalls or other relevant overhead with starting issuing the next warp.
    
    \item[4)] By analyzing the latency and throughput under ILP = 1 with different \#warp, we observe that instructions from different warps can run in the same sub-core Tensor Cores computation pipeline. Specifically, increasing \#warps from 4 to 8 (4$\times$2) and 12 (4$\times$3) will only lead to one extra cycle latency, but the latency increase significantly if \#warp goes to 16. This is because that 12 warp with ILP = 1 is equivalent to four warps with ILP = 3 from the workload perspective. Based on our previous observation in \textbf{3)}, four warps with ILP = 3 can achieve the full Tensor Cores usage and increasing ILP will not lead to higher throughput but significantly longer latency. 
    

    \item[5)] \#warps = 6 is a special case. First, the latency of \#warps = 6 and \#warps = 8 are always the same under different ILPs. Second, for ILP $\geq$ 3, the throughput drops if increasing \#warps from 4 to 6. This is because when there is 6 warps resident in an SM, the first four warps will be issued to the Tensor Cores computation pipeline. Since the ILP = 3 is large enough to fulfill the whole Tensor Cores computation pipeline, the second two warps can not be issued until there are available resources freed by the first four warps. When the second two warps start to execute on two sub-cores, the other two sub-cores are idle which causes the drop in the overall throughput. 
    
    \item[6)] In general, to reach the peak performance when programming Tensor Cores using mma.m16n8k16 instruction,
    \#warp should be at least four and ideally a multiple of 4. Although four warps with sufficient ILP (i.e. $\geq$ 3 ) can achieve near peak performance, eight warps with ILP $\geq$ 2 should be used whenever possible 
    
\end{itemize}

\begin{figure}[htp]
     \centering
     \begin{subfigure}[*]{0.48\textwidth}
         \centering
         \includegraphics[width=\textwidth]{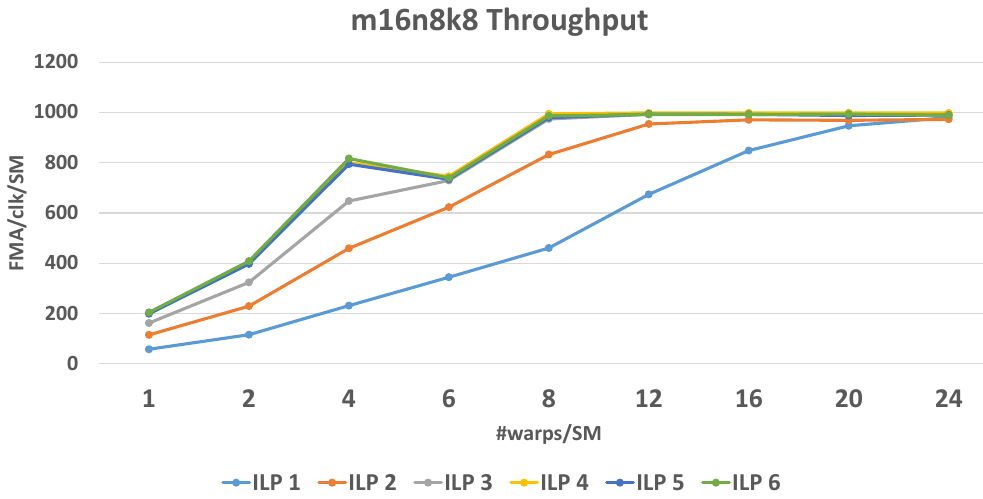}
     \end{subfigure}
     
    \begin{subfigure}[*]{0.48\textwidth}
         \centering
         \includegraphics[width=\textwidth]{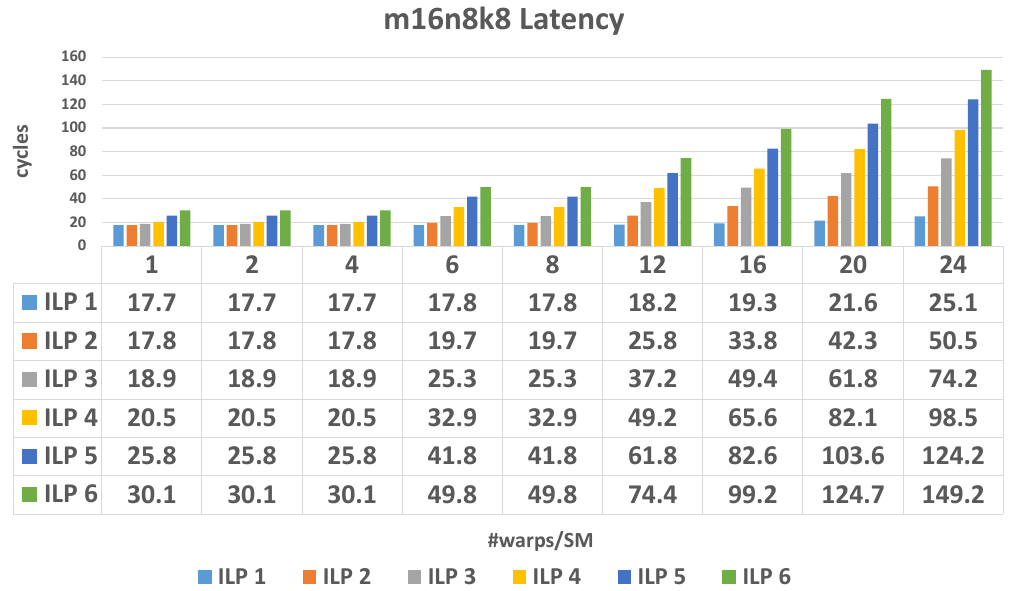}
     \end{subfigure}
\caption{Throughput and latency of mma.m16n8k8 instruction under different settings on A100 Tensor Cores.}
\label{fig:A100-m16n8k8}
\end{figure}


Figure~\ref{fig:A100-m16n8k8} presents the results of m16n8k8 instruction. Following the analysis of previous m16n8k16 instruction. We provide additional observations as follows:
\begin{itemize}
    \item[7)] The instruction completion latency is around 18 cycles and the peak throughput is around 1000 FMA/clk/SM.
    \item[8)] Like the observations in \textbf{3)}, there are two convergence points -- \#warp = 4, ILP = 4 and \#warp = 8, ILP = 3. However, the throughput gap between these two points is more significant (800 v.s. 1000). This suggests that the intra-warp synchronization stalls overhead of mma.m16n8k8 are significantly higher than previously discussed mma.m16n8k16 (i.e. 900 v.s. 1000). Therefore, when choosing mma.m16n8k8 to program Tensor Cores, at least eight warps should be allocated to achieve ideal performance.
\end{itemize}

\begin{table*}[htp]
\centering
\caption{\textcolor{Highlight}{mma instructions with different data types on A100 Tensor Cores. The peak performance of FP16, TF32, INT8, INT4 and Binary are 1024, 512, 2048, 4096, and 16384 FMA/clk/SM respectively~\cite{AmpereWhite}}. }
\label{table:A100-mma}
  \centering
          \small
        \setlength\tabcolsep{2pt}
\begin{tabular}{|c|c|c|c|c|c|c|c|c|c|}
\hline
A/B  & C/D   & Shape    & \begin{tabular}[c]{@{}c@{}}Completion Latency\\ (cycles)\end{tabular} & (\#warp,ILP) & \begin{tabular}[c]{@{}c@{}}Latency\\ (cycles)\end{tabular} & \begin{tabular}[c]{@{}c@{}}Throughput\\ (FMA/clk/SM)\end{tabular} & (\#warp,ILP) & \begin{tabular}[c]{@{}c@{}}Latency\\ (cycles)\end{tabular} & \begin{tabular}[c]{@{}c@{}}Throughput\\ (FMA/clk/SM)\end{tabular} \\ \hline
FP16 & FP32  & m16n8k16 & 24.7                                                                  & 4,3          & 27.4                                                       & 897.6                                                             & 8,2          & 32.6                                                       & 1004.2                                                            \\ \hline
FP16 & FP32  & m16n8k8  & 17.7                                                                  & 4,4          & 20.5                                                       & 800.2                                                             & 8,3          & 25.3                                                       & 974.1                                                             \\ \hline
FP16 & FP16  & m16n8k16 & 24.4                                                                  & 4,3          & 27.1                                                       & 907.1                                                             & 8,2          & 32.9                                                       & 996.6                                                             \\ \hline
FP16 & FP16  & m16n8k8  & 17.7                                                                  & 4,4          & 19.1                                                       & 860.9                                                             & 8,3          & 24.5                                                       & 1002.6                                                            \\ \hline
TF32 & FP32  & m16n8k8  & 25                                                                    & 4,3          & 28.2                                                       & 435.9                                                             & 8,2          & 33.3                                                       & 492.4                                                             \\ \hline
TF32 & FP32  & m16n8k4  & 18.1                                                                  & 4,4          & 20.9                                                       & 392.6                                                             & 8,3          & 25.7                                                       & 477.5                                                             \\ \hline
INT8 & INT32 & m8n8k16  & 15.9                                                                  & 4,4          & 20.1                                                       & 813.2                                                             & 8,2          & 16.4                                                       & 998.3                                                             \\ \hline
INT8 & INT32 & m16n8k32 & 24.7                                                                  & 4,3          & 27.1                                                       & 1812.4                                                            & 8,2          & 32.9                                                       & 1986.5                                                            \\ \hline
INT8 & INT32 & m16n8k16 & 17.6                                                                  & 4,4          & 20.9                                                       & 1570.1                                                            & 8,3          & 25.1                                                       & 1965.1                                                            \\ \hline
\textcolor{Highlight}{INT4} & \textcolor{Highlight}{INT32} & m16n8k32 &18.1                                                                  & 4,4          & 22.1                                                       & 2971.1                                                            & 8,3          & 27.1                                                       & 3630.0                                                            \\ \hline
\textcolor{Highlight}{INT4} & \textcolor{Highlight}{INT32} & m16n8k64 &26.1                                                                  & 4,3          & 28.1                                                       & 3497.9                                                            & 8,2          & 35.8                                                       & 3660.8                                                            \\ \hline
\textcolor{Highlight}{Binary} & \textcolor{Highlight}{INT32} & m16n8k128 &18.1                                                                  & 4,4          & 22.1                                                       & 11884.3                                                            & 8,3          & 27.1                                                       & 14515.1                                                            \\ \hline
\textcolor{Highlight}{Binary} & \textcolor{Highlight}{INT32} & m16n8k256 &26.0                                                                  & 4,3          & 28.1                                                       & 13985.4                                                            & 8,2          & 35.8                                                       & 14643.4                                                            \\ \hline
\end{tabular}
\end{table*}

\begin{table*}[htp]
\centering
\caption{\textcolor{Highlight}{Tensor Cores performance of RTX3070Ti Ampere GPU~\cite{GA102}. Unlike A100 which is designed for high-end servers. RTX3070Ti is for graphical applications and there are noticeable differences between these two GPUs}.}
\label{table:RTX3070Ti-mma}
\centering
\small
\setlength\tabcolsep{2pt}
\begin{tabular}{|c|c|c|c|c|c|c|c|c|c|}
\hline
A/B  & C/D   & Shape    & \begin{tabular}[c]{@{}c@{}}Completion Latency\\ (cycles)\end{tabular} & (\#warp,ILP) & \begin{tabular}[c]{@{}c@{}}Latency\\ (cycles)\end{tabular} & \begin{tabular}[c]{@{}c@{}}Throughput\\ (FMA/clk/SM)\end{tabular} & (\#warp,ILP) & \begin{tabular}[c]{@{}c@{}}Latency\\ (cycles)\end{tabular} & \begin{tabular}[c]{@{}c@{}}Throughput\\ (FMA/clk/SM)\end{tabular} \\ \hline
FP16 & FP32  & m16n8k16 & 33                                                                    & 4,1          & 33                                                         & 248.2                                                             & 8,1          & 64.8                                                       & 252.7                                                             \\ \hline
FP16 & FP32  & m16n8k8  & 18.8                                                                  & 4,2          & 32.3                                                       & 253.9                                                             & 8,1          & 32.4                                                       & 253.2                                                             \\ \hline
FP16 & FP16  & m16n8k16 & 24                                                                    & 4,2          & 32.2                                                       & 509.4                                                             & 8,1          & 32.3                                                       & 506.9                                                             \\ \hline
FP16 & FP16  & m16n8k8  & 17.7                                                                  & 4,3          & 24                                                         & 511.8                                                             & 8,2          & 32.3                                                       & 507.8                                                             \\ \hline
TF32 & FP32  & m16n8k8  & 33.3                                                                  & 4,1          & 33.4                                                       & 122.6                                                             & 8,1          & 64.6                                                       & 126.8                                                             \\ \hline
TF32 & FP32  & m16n8k4  & 19.1                                                                  & 4,2          & 32.7                                                       & 125.3                                                             & 8,1          & 32.6                                                       & 125.7                                                             \\ \hline
INT8 & INT32 & m8n8k16  & 15.9                                                                  & 4,4          & 19.3                                                       & 848.9                                                             & 8,2          & 16.2                                                       & 1008.5                                                            \\ \hline
INT8 & INT32 & m16n8k32 & 24.3                                                                  & 4,2          & 32.2                                                       & 1017.2                                                            & 8,1          & 32.1                                                       & 1023.2                                                            \\ \hline
INT8 & INT32 & m16n8k16 & 17.7                                                                  & 4,3          & 24.1                                                       & 1018.2                                                            & 8,2          & 32.6                                                       & 1005.4                                                            \\ \hline

\textcolor{Highlight}{INT4} & \textcolor{Highlight}{INT32} & m16n8k32 & 17.3                                                                  & 4,3          & 24.9                                                       & 1967.9                                                            & 8,2          & 32.3                                                       & 2031.7                                                            \\ \hline
\textcolor{Highlight}{INT4} & \textcolor{Highlight}{INT32} & m16n8k64 & 24.5                                                                  & 4,2          & 33.3                                                       & 1967.9                                                            & 8,1          & 32.5                                                      & 2013.5                                                            \\ \hline
\textcolor{Highlight}{Binary} & \textcolor{Highlight}{INT32} & m16n8k128 & 17.3                                                                  & 4,3          & 24.8                                                       & 7908.3                                                            & 8,2          & 32.3                                                      & 8127.2                                                            \\ \hline
\textcolor{Highlight}{Binary} & \textcolor{Highlight}{INT32} & m16n8k256 & 24.6                                                                  & 4,2          & 33.3                                                       & 7871.9                                                            & 8,1          & 32.5                                                      & 8053.9                                                            \\ \hline
\end{tabular}
\end{table*}

\begin{table*}[htp]
\centering
\caption{Tensor Cores performance of RTX2080Ti Turing GPU~\cite{TuringWhite}.}
\label{table:RTX2080Ti-mma}
\centering
\small
\setlength\tabcolsep{2pt}
\begin{tabular}{|c|c|c|c|c|c|c|c|c|c|}
\hline
A/B  & C/D   & Shape   & \begin{tabular}[c]{@{}c@{}}Completion Latency\\ (cycles)\end{tabular} & (\#warp,ILP) & \begin{tabular}[c]{@{}c@{}}Latency\\ (cycles)\end{tabular} & \begin{tabular}[c]{@{}c@{}}Throughput\\ (FMA/clk/SM)\end{tabular} & (\#warp,ILP) & \begin{tabular}[c]{@{}c@{}}Latency\\ (cycles)\end{tabular} & \begin{tabular}[c]{@{}c@{}}Throughput\\ (FMA/clk/SM)\end{tabular} \\ \hline
FP16 & FP32  & m16n8k8 & 17.3                                                                  & 4,2          & 32.5                                                       & 252.4                                                             & 8,1          & 32.1                                                       & 255.1                                                             \\ \hline
FP16 & FP16  & m16n8k8 & 14.7                                                                  & 4,2          & 17.5                                                       & 467.9                                                             & 8,1          & 16.1                                                       & 509.4                                                             \\ \hline
INT8 & INT32 & m8n8k16 & 11                                                                    & 4,3          & 14.5                                                       & 846.1                                                             & 8,2          & 16.2                                                       & 1012.6                                                            \\ \hline
\end{tabular}
\end{table*}


Table~\ref{table:A100-mma} lists the rest of the data types supported in Ampere Tensor Cores. We summarize their completion latency and two key convergence points, one with 4 warps and the other one with 8 warps. Note that all instructions can achieve near peak performance except for the INT8 data type with m8n8k16 shape, which can only achieve near half peak performance. By comparing the performance of this instruction on Turing Tensor Cores (Table~\ref{table:RTX2080Ti-mma}), which can achieve near peak performance. We find the reason can be that m8n8k16 is an old shape optimized for Turing Tensor Cores, the Ampere Tensor Cores prefer new shapes with m16n8 (i.e. m16n8k16 and m16n8k32).

Table~\ref{table:RTX3070Ti-mma} summarizes the dense Tensor Cores performance on an RTX3070Ti Ampere GPU which is designed for video gaming applications. Compared to high-end data center A100 GPU, there are two key differences: First, the Tensor Cores in RTX3070Ti are less powerful and offer less peak throughput for all data types. Second, when using FP32 as the data type of matrix C and D, the performance is half of that of using FP16 as the data type. On the other hand, the data type of matrix C/D will not affect the peak performance of the A100 Tensor Cores. The differences suggest a careful check and adjustment is necessary when using Tensor Cores to accelerate gaming or video applications, especially if the original program is developed on a data center A100 GPU, since the difference may lead to undesired performance. 

Table~\ref{table:RTX2080Ti-mma} presents the Tensor Cores performance on a RTX2080 Turing GPU. The Turing Tensor Cores is the predecessor of the Ampere Tensor Cores, it supports fewer instruction shapes and data types. It is interesting to note that the Dense FMA latency of Ampere Tensor Cores does not improve compared to Turing Tensor Cores. For instance, the latency of mma.m16n8k8 is 17.3 cycles on RTX2080Ti Turing Tensor Cores, which is close to the number 17.7 cycles in A100 Ampere Tensor Cores. 

To summarize, in this Section, we experimented on different mma PTX instructions on Ampere and Turing Tensor Cores. Since Volta Tensor Cores only supports legacy programming interfaces and has been extensively studied by previous work as explained in Section~\ref{sec:related-work}, we do not include the experiments of Volta GPU Tensor Cores. Based on our experiments, we indicate how peak performance can be achieved (i.e. i.e. how to choose proper \#warps and ILP).

\section{Sparse FMA}\label{sect:SparseTC}
Sparse matrix multiplication (SpMM) is an important computation pattern in Deep Learning domains. Plenty of efforts have been invested in optimizing SpMM software on GPUs over the past years~\cite{niu2022tilespgemm,anh2016balanced,chen2018performanceSP,dalton2015optimizing,deveci2018multithreaded,kunchum2017improving,lee2020optimization,parger2020speck,winter2019adaptive,xie2021pattern}.
Although significant progress has been made, the performance of SpMM software on GPUs is not satisfied in terms of the hardware throughput bound. On the other hand, due to the irregularity of non-zeros in sparse matrix, it is extremely challenging to adjust GPGPU architecture to accelerate SpMM via hardware approach with acceptable overhead. Recently, fine-grained structured sparsity, 
especially N:M sparsity (i.e.  there are N non-zeros for every M consecutive elements), has shown a promising path to compress Deep Learning Models while maintaining a hardware-friendly sparse pattern ~\cite{sun2021dominosearch,zhou2021learning,hubara2021accelerated}. Ampere Tensor Cores are the first general purpose hardware empowered with this kind of fine-grained N:M sparse hardware acceleration.


\begin{figure}[htp]
    \centering
    \includegraphics[width=8.8cm]{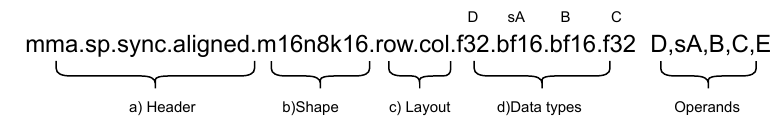}
    \caption{The mma.sp instruction performs D = sA$\times$B + C sparse-dense matrix multiplication. sA contains the non-zero values of fine-grained 2:4 sparse matrix A (m$\times$k) and the shape of sA, therefore, is m$\times$k/2. Matrix B is still a dense matrix with the shape k$\times$n. The final operand e is the sparse metadata containing the indexes of Matrix A.} 
    \label{fig:sparseInst}
\end{figure}

\begin{figure}[htp]
    \centering
    \includegraphics[width=8cm]{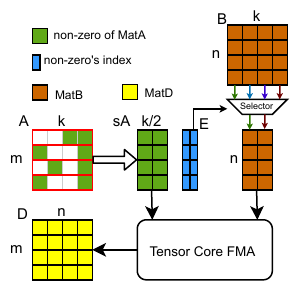}
    \caption{Illustration of fine-grained 2:4 SpMM~\cite{AmpereWhite}. Matrix C is not shown for simplification.}
    \label{fig:sparse}
\end{figure}

Unlike dense Tensor Cores which can be programmed by both modern mma instruction and legacy wmma.mma instruction, the sparse computation can only be exploited by modern mma.sp instruction. Figure~\ref{fig:sparseInst} shows an example of sparse instruction for BF16 computation. Figure~\ref{fig:sparse} illustrates 2:4 sparse Tensor Cores computation. To use the sparse Tensor Cores correctly, programmers need to first compress the sparse matrix A into the non-zeros matrix (denoted as sA) with the index metadata. The sparse pattern of matrix A should strictly follow the 2:4 fine-grained sparsity which means for every four consecutive elements along the K dimension, there are two non-zero values. Therefore, after compression, the shape of matrix A will become m$\times$k/2. The index metadata encodes the relative location of every non-zero in 2 bits because 2 bits can represent all possibilities of the four consecutive elements (00,01,10,11). On the other hand, matrix B is still stored in the original dense format, the hardware selector determines which values should participate in the computation on the fly based on the indexes provided by metadata. 


\subsection*{Experimental results}

\begin{figure}[htp]
     \centering
     \begin{subfigure}[*]{0.48\textwidth}
         \centering
         \includegraphics[width=\textwidth]{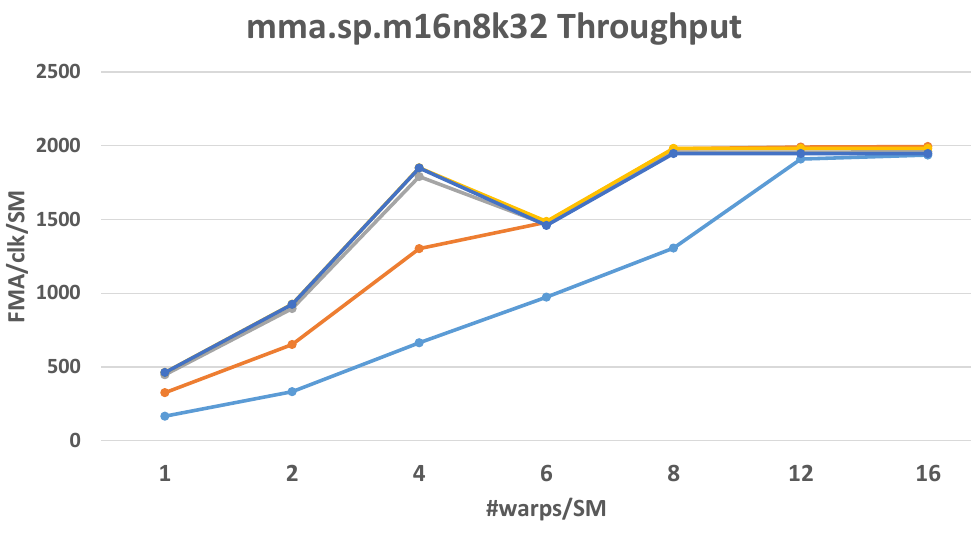}
     \end{subfigure}
    \begin{subfigure}[*]{0.48\textwidth}
         \centering
         \includegraphics[width=\textwidth]{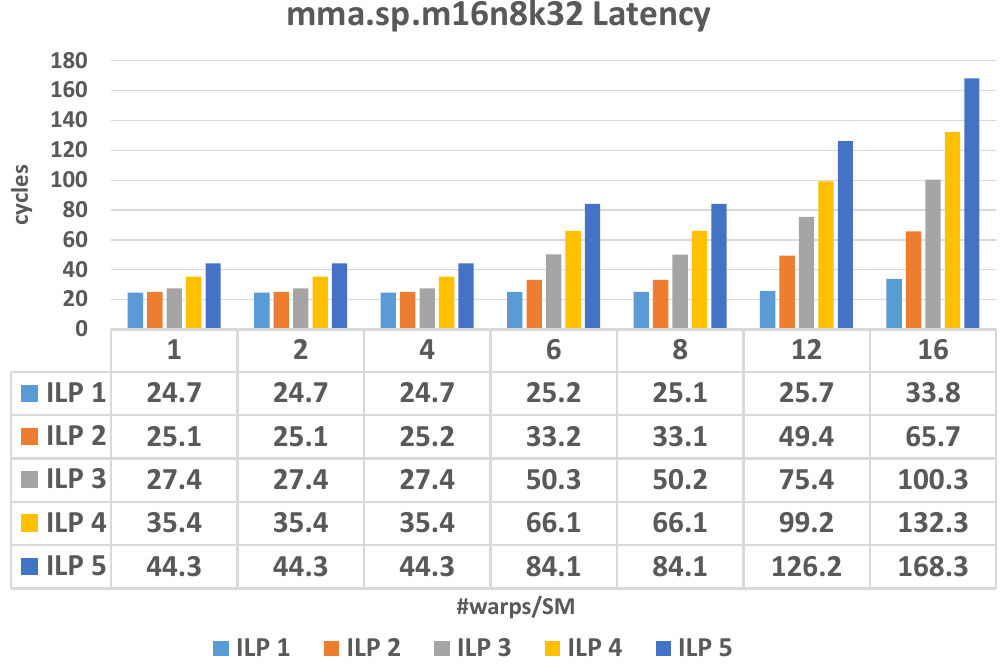}
     \end{subfigure}
\caption{Throughput and latency of mma.sp.m16n8k32 instruction under different settings on A100 Tensor Cores.}
\label{fig:sp-m16n8k32}
\end{figure}


Figure~\ref{fig:sp-m16n8k32} presents the latency and throughput of sparse instruction with shape m16n8k32 instruction. Note there are some interesting similarities between sparse m16n8k32 and dense m16n8k16 (Fig.~\ref{fig:A100-m16n8k16}). We provide the following observations based on the experimental results:

\begin{itemize}
    \item[1)]  The completion latency is 24.7 cycles which is the same as its dense mma counterpart (mma.m16n8k16). Furthermore, the latency results under different settings are also almost the same as the results of dense mma.m16n8k16 if we consider the experimental deviations. This observation indicates that the selector for matrix B is integrated with the Tensor Cores hardware pipeline. Although dense FMA operation does not need the selector, it will go through the selector as sparse FMA operation which therefore results in the same execution cycles.
    
    \item[2)] The peak throughput achieves at 2000 which is 2x higher than the dense one. The throughput graph shows the same trend as the dense one, the only difference is that the numbers on Y-axis double. This indicates that the improvement comes from skipping the zero multiplication. It confirms that a sparse feature can improve the throughput but can not reduce the latency.
\end{itemize}

\begin{figure}[htp]
     \centering
     \begin{subfigure}[*]{0.48\textwidth}
         \centering
         \includegraphics[width=\textwidth]{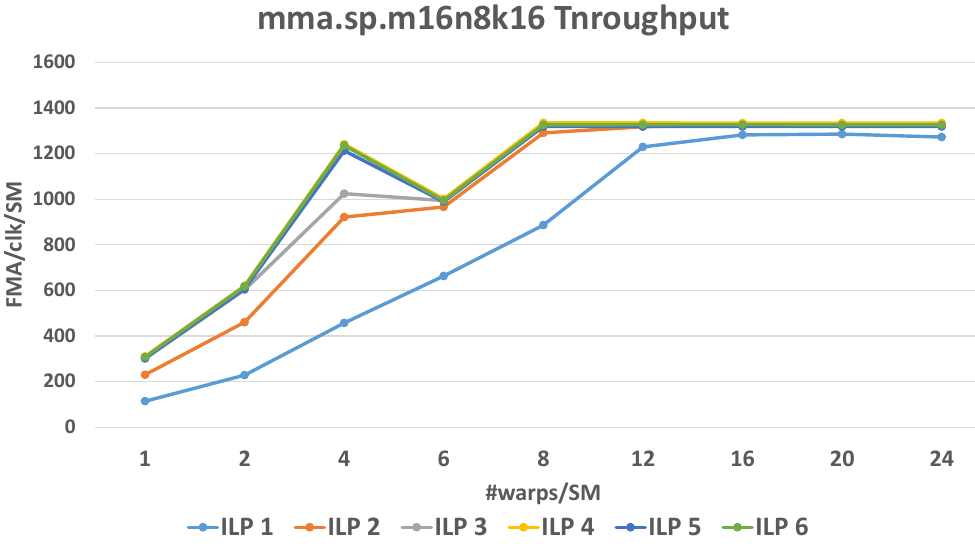}
     \end{subfigure}
     
    \begin{subfigure}[*]{0.48\textwidth}
         \centering
         \includegraphics[width=\textwidth]{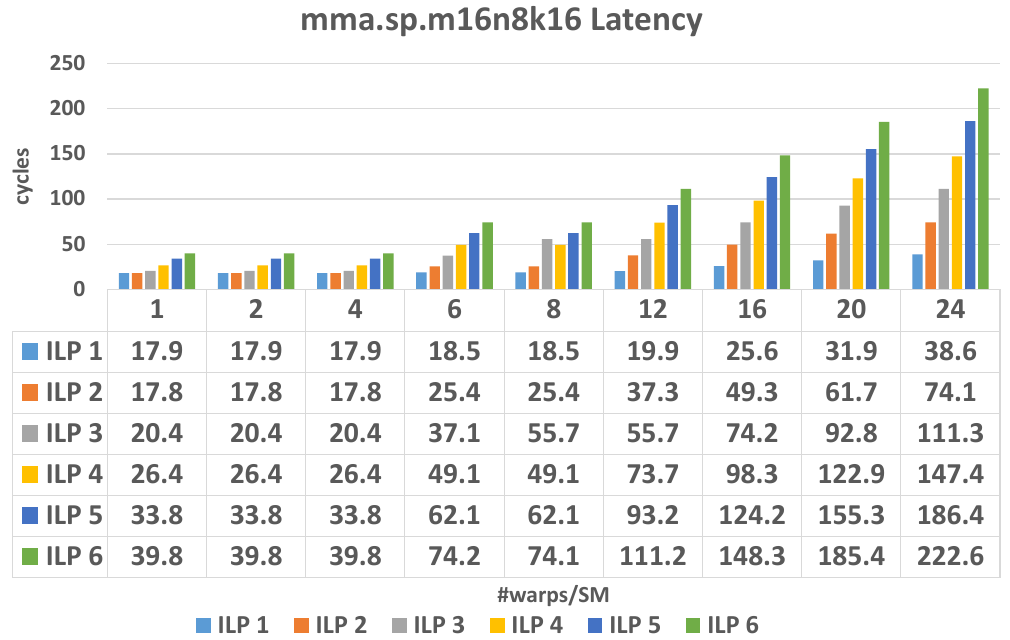}
     \end{subfigure}
\caption{Throughput and latency of mma.sp.m16n8k16 instruction under different settings on A100 Tensor Cores.}
\label{fig:sp-m16n8k16}
\end{figure}

Figure~\ref{fig:sp-m16n8k16} presents the results of mma.sp.m16n8k16. According to the discussions of mma.sp.m16n8k32, the behaviors of mma.sp.m16n8k16 should show similarities with dense mma.m16n8k8. However, we notice that the experimental results are not as expected: 

\begin{itemize}
    \item[3)] The instruction completion latency is 17.9 cycles which are close to the latency of dense m16n8k8 (i.e., 17.7 cycles).
    \item[4)] The throughput behaviors are not as expected. The peak throughput can only reach 1300. There is a significant gap compared to theoretical peak performance 2000. Furthermore, measured throughput under different settings is also significantly lower than the expected values (i.e. twice the corresponding dense throughput). 
    
\end{itemize}

\begin{table*}[htp]
\centering
\caption{Sparse Tensor Cores performance of A100 Ampere GPU.}
\label{table:A100-mmasp}
\centering
\small
\setlength\tabcolsep{2pt}
\begin{tabular}{|c|c|c|c|c|c|c|c|c|c|}
\hline
A/B  & C/D   & Shape    & \begin{tabular}[c]{@{}c@{}}Completion Latency\\ (cycles)\end{tabular} & (\#warp,ILP) & \begin{tabular}[c]{@{}c@{}}Latency\\ (cycles)\end{tabular} & \begin{tabular}[c]{@{}c@{}}Throughput\\ (FMA/clk/SM)\end{tabular} & (\#warp,ILP) & \begin{tabular}[c]{@{}c@{}}Latency\\ (cycles)\end{tabular} & \begin{tabular}[c]{@{}c@{}}Throughput\\ (FMA/clk/SM)\end{tabular} \\ \hline
FP16 & FP32  & m16n8k32 & 24.7                                                                  & 4,3          & 27.4                                                       & 1791.9                                                            & 8,2          & 33.1                                                       & 1979.1                                                            \\ \hline
FP16 & FP32  & m16n8k16 & 17.8                                                                  & 4,3          & 20.4                                                       & 1024.5                                                            & 8,2          & 25.4                                                       & 1290.5                                                            \\ \hline
FP16 & FP16  & m16n8k32 & 24.3                                                                  & 4,3          & 26.6                                                       & 1850.9                                                            & 8,2          & 32.4                                                       & 2019.8                                                            \\ \hline
FP16 & FP16  & m16n8k16 & 17.6                                                                  & 4,3          & 19.8                                                       & 1242.9                                                            & 8,2          & 24.9                                                       & 1318.2                                                            \\ \hline
TF32 & FP32  & m16n8k16 & 24.9                                                                  & 4,3          & 28.3                                                       & 868.2                                                             & 8,2          & 33.9                                                       & 981.2                                                             \\ \hline
TF32 & FP32  & m16n8k8  & 18.2                                                                  & 4,3          & 20.6                                                       & 597.8                                                             & 8,2          & 25.5                                                       & 643.6                                                             \\ \hline
INT8 & INT32 & m16n8k64 & 24.7                                                                  & 4,3          & 27.7                                                       & 3544.7                                                            & 8,2          & 33.1                                                       & 3961.5                                                            \\ \hline
INT8 & INT32 & m16n8k32 & 17.9                                                                  & 4,3          & 20.4                                                       & 2403.9                                                            & 8,2          & 25.4                                                       & 2665.2                                                            \\ \hline

\end{tabular}
\end{table*}

\begin{table*}[htp]
\centering
\caption{Sparse Tensor Cores performance of RTX3070Ti Ampere GPU.}
\label{table:RTX3070Ti-mmasp}
\centering
\small
\setlength\tabcolsep{2pt}
\begin{tabular}{|c|c|c|c|c|c|c|c|c|c|}
\hline
A/B  & C/D   & Shape    & \begin{tabular}[c]{@{}c@{}}Completion Latency\\ (cycles)\end{tabular} & (\#warp,ILP) & \begin{tabular}[c]{@{}c@{}}Latency\\ (cycles)\end{tabular} & \begin{tabular}[c]{@{}c@{}}Throughput\\ (FMA/clk/SM)\end{tabular} & (\#warp,ILP) & \begin{tabular}[c]{@{}c@{}}Latency\\ (cycles)\end{tabular} & \begin{tabular}[c]{@{}c@{}}Throughput\\ (FMA/clk/SM)\end{tabular} \\ \hline
FP16 & FP32  & m16n8k32 & 33                                                                    & 4,1          & 33                                                         & 496.5                                                             & 8,1          & 64.1                                                       & 511.2                                                             \\ \hline
FP16 & FP32  & m16n8k16 & 18.8                                                                  & 4,2          & 32.3                                                       & 507.8                                                             & 8,1          & 32.4                                                       & 506.2                                                             \\ \hline
FP16 & FP16  & m16n8k32 & 24.3                                                                  & 4,2          & 32                                                         & 1022.2                                                            & 8,1          & 32.1                                                       & 1022.3                                                            \\ \hline
FP16 & FP16  & m16n8k16 & 17.7                                                                  & 4,3          & 24.2                                                       & 1013.4                                                            & 8,2          & 32                                                         & 1023.1                                                            \\ \hline
TF32 & FP32  & m16n8k16 & 33.2                                                                  & 4,1          & 33.2                                                       & 247                                                               & 8,1          & 64.2                                                       & 255.1                                                             \\ \hline
TF32 & FP32  & m16n8k8  & 19                                                                    & 4,2          & 32.5                                                       & 252.5                                                             & 8,1          & 32.4                                                       & 253.2                                                             \\ \hline
INT8 & INT32 & m16n8k64 & 24.3                                                                  & 4,2          & 64.2                                                       & 2040.2                                                            & 8,1          & 32.1                                                       & 2039.5                                                            \\ \hline
INT8 & INT32 & m16n8k32 & 17.7                                                                  & 4,3          & 24.2                                                       & 2028.8                                                            & 8,2          & 32.3                                                       & 2031.8                                                            \\ \hline
\end{tabular}
\end{table*}

Table~\ref{table:A100-mmasp} lists sparse FMA instructions for the rest of the data types and shapes on A100 Tensor Cores. For each data type, there are two supported shapes with different k sizes. Like the behavior of the BF16 data type in Fig.~\ref{fig:sp-m16n8k16}, sparse FMA instructions with the smaller k can not reach the theoretical peak performance. To study whether this is a special issue for A100 GPU or a general problem for Ampere Tensor Cores, we check the results on an RTX3070Ti Ampere GPU as presented in Table~\ref{table:RTX3070Ti-mmasp}. Interestingly, the sparse Tensor Cores performance of RTX3070Ti does not show the same issue and the instruction with a smaller k can also reach the same throughput as the instruction with a larger k.   

To summarize, the sparse Tensor Cores in general can provide twice throughput improvement compared to the dense Tensor Cores but can not reduce the completion latency. However, programmers and users must be careful when deciding which PTX instruction should be used especially when the targeted platform is the data-center A100 GPUs. The instruction with a smaller k can not provide the expected peak throughput. However, the vendor does not document the reason for this issue.



\section{Data movement} \label{sect:DataMovement}

In this Section, we study the data movement instructions related to Tensor Cores. As introduced in Section~\ref{sec:Background}, before using Tensor Cores instructions (mma and mma.sp), the input data should be first loaded from Shared Memory to the Register File via data movement instructions. This can be done by special Tensor Cores instructions wmma.load and ldmatrix or generic ld.shared instruction.

\begin{figure}[htp]
    \centering
    \includegraphics[width=8cm]{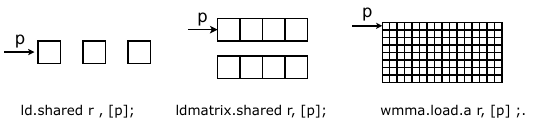}
    \caption{Differences of three data movement instructions.}
    \label{fig:ld_gra}
\end{figure}

Figure~\ref{fig:ld_gra} illustrates the differences of these three instructions. The generic per-thread ld.shared instruction receives a pointer (p) of an element in Shared Memory and loads it to the destination register (r). For each warp, 32 elements will be loaded from the Shared Memory and each thread holds one register to store the element. On the other hand, wmma.load and ldmatrix run in the per-warp scheme so the 32 threads of a specific warp will cooperatively complete the job. For the legacy wmma.load instruction, there is only one address pointer p to the starting element of the matrix. The elements of the matrix will then be loaded to one or more registers of each thread within the warp depending on the loading shape specified by the instruction. For example, when using wmma.load.a.m16n16k16.FP16, a warp loads matrix A with 16$\times$16 FP16 elements from the Shared Memory. Then each thread will use 16$\times$16/32 = 8 registers to store the elements. 

\begin{figure}[htp]
    \centering
    \includegraphics[width=7.5cm]{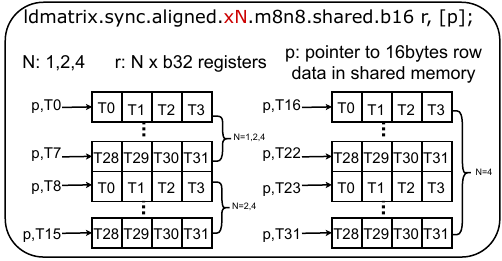}
    \caption{ Functionality overview. p is the source operand and is provided by threads. If N = 4, then every thread will provide a valid p. if N $<$ 4, partial threads (e.g. T0 - T7 for N = 1) will provide a valid p, and other threads' p will be ignored. }
    \label{fig:func_ldmatrix}
\end{figure}

The new ldmatrix instruction is a bit tricky. It runs in a per-warp scheme like wmma.load, but it offers more flexibility. Figure~\ref{fig:func_ldmatrix} shows the functionality of ldmatrix instruction. It loads N$\times$128 bytes per warp cooperatively from the Shared Memory and stores it in the Register File where each thread stores N$\times$4 bytes. Compared to generic per-thread ld.shared instruction, ldmatrix is less flexible because it requires a group of four consecutive threads to load 16 consecutive bytes (as a group). However, ldmatrix requires fewer source operands (each p will be broadcast to four threads), and the data layout eventually fetched to the Register File matches the arrangement of the inputs to the Tensor Cores mma/mma.sp instructions~\cite{NVPTX}. Compared to legacy per-warp wmma.load instructions, ldmatrix is more flexible because the 16-bytes groups do not need to be consecutive, while wmma.load requires the whole matrix stored consecutively. By contrast, additional index computations are necessary if using generic ld.shared instruction, which will cause extra overhead. The flexibility of ldmatrix enables using some special data layouts to avoid bank conflict such as permuted data layout used in CUTLASS~\cite{CUTLASS}. Furthermore, ldmatrix can be used to load matrix A, matrix B, and matrix C, while wmma.load has dedicated instructions for different operands as shown in Figure~\ref{fig:wmma.load}. \textcolor{Highlight}{We present ablation experiments in Appendix~\ref{appendix:flexibilityOfLdmatrix} to demonstrate how the flexibility of ldmatrix can be exploited to improve performance.}



\begin{figure}[htp]
    \centering
    \includegraphics[width=7.5cm]{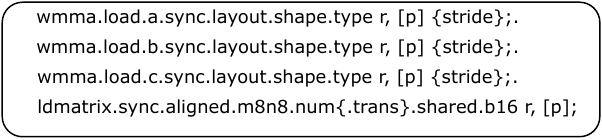}
    \caption{{wmma.load and ldmatrix~\cite{NVPTX}}. ldmatrix can be used for loading a,b, and c operands but wmma has dedicated instruction for each of the operands.}
    \label{fig:wmma.load}
\end{figure}




Unlike mma and mma.sp instructions which are designed for the novel Tensor Cores hardware, ldmatrix instruction is an alternative instruction for using the existing hardware (i.e. loading data from the Shared Memory to the Register File). Therefore, the previous study of GPU Shared Memory~\cite{DissectMemory} also applies to the ldmatrix instruction. For instance, the Shared Memory of modern NVIDIA GPUs (e.g. Ampere, Turing, Volta) has 32 banks and the width of each bank is 4 bytes which give 32$\times$4 = 128 bytes/clk as the theoretical bandwidth. This number is also the bandwidth bound of ldmatrix instruction.

\begin{table}[htp]
\centering
\caption{The loading bytes per instruction of ldmatrix and ld.shared.}
\label{table:ldmatrix-ld.shared}
\begin{tabular}{|c|c|c|}
\hline
              & bytes/warp & bytes/thread \\ \hline
ldmatrix.x1   & 128        & 4            \\ \hline
ldmatrix.x2   & 256        & 8            \\ \hline
ldmatrix.x4   & 512        & 16           \\ \hline
ld.shared.u32 & 128        & 4            \\ \hline
ld.shared.u64 & 256        & 8            \\ \hline
\end{tabular}
\end{table}

Table~\ref{table:ldmatrix-ld.shared} lists the instruction workload of ldmatrix and ld.shared. Since traditional ld.shared offers better granularity, we microbenchmark both ldmatrix and ld.shared and make comparisons for better understanding. 

\subsection*{Experimental results}

\begin{figure}[htp]
     \centering
     \begin{subfigure}[*]{0.48\textwidth}
         \centering
         \includegraphics[width=\textwidth]{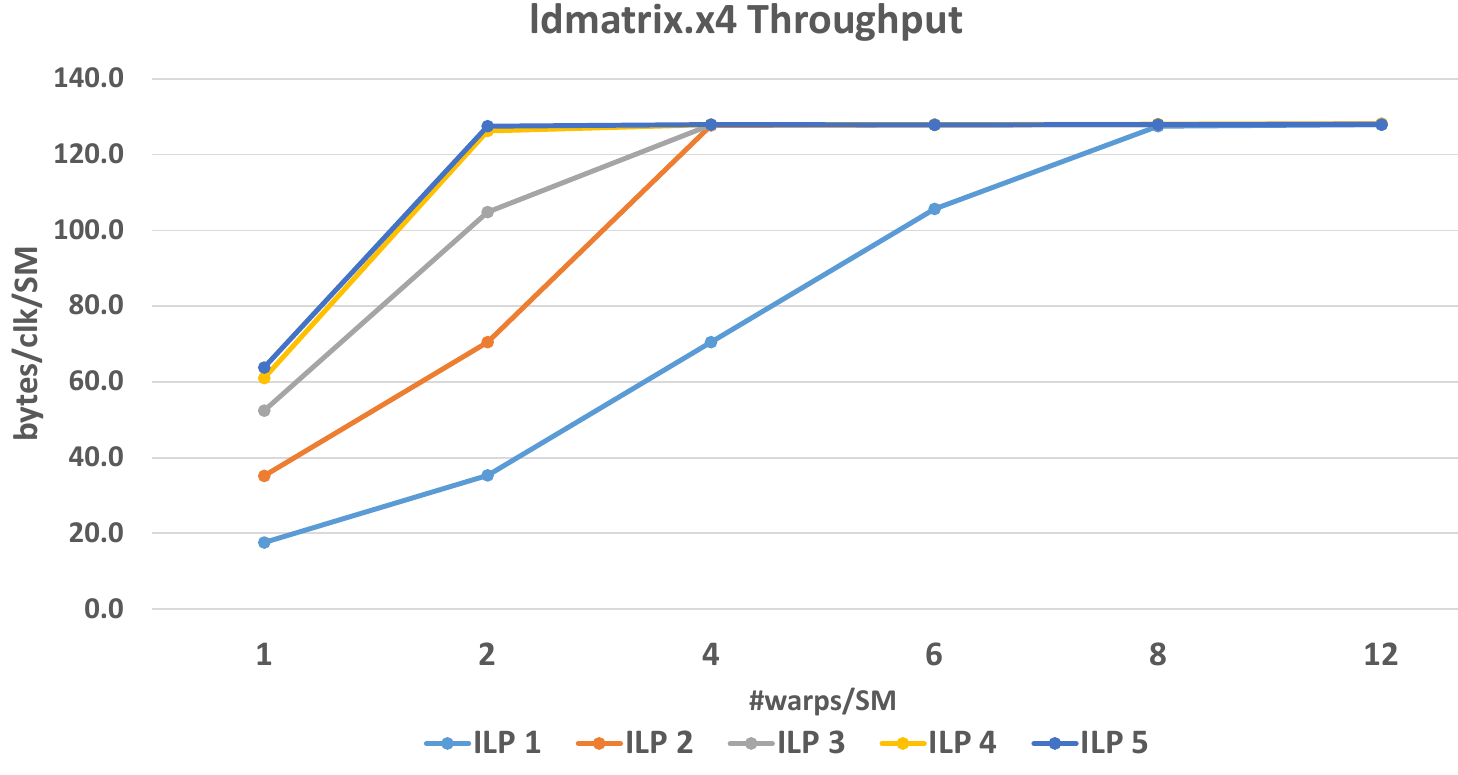}
     \end{subfigure}
    \begin{subfigure}[*]{0.48\textwidth}
         \centering
         \includegraphics[width=\textwidth]{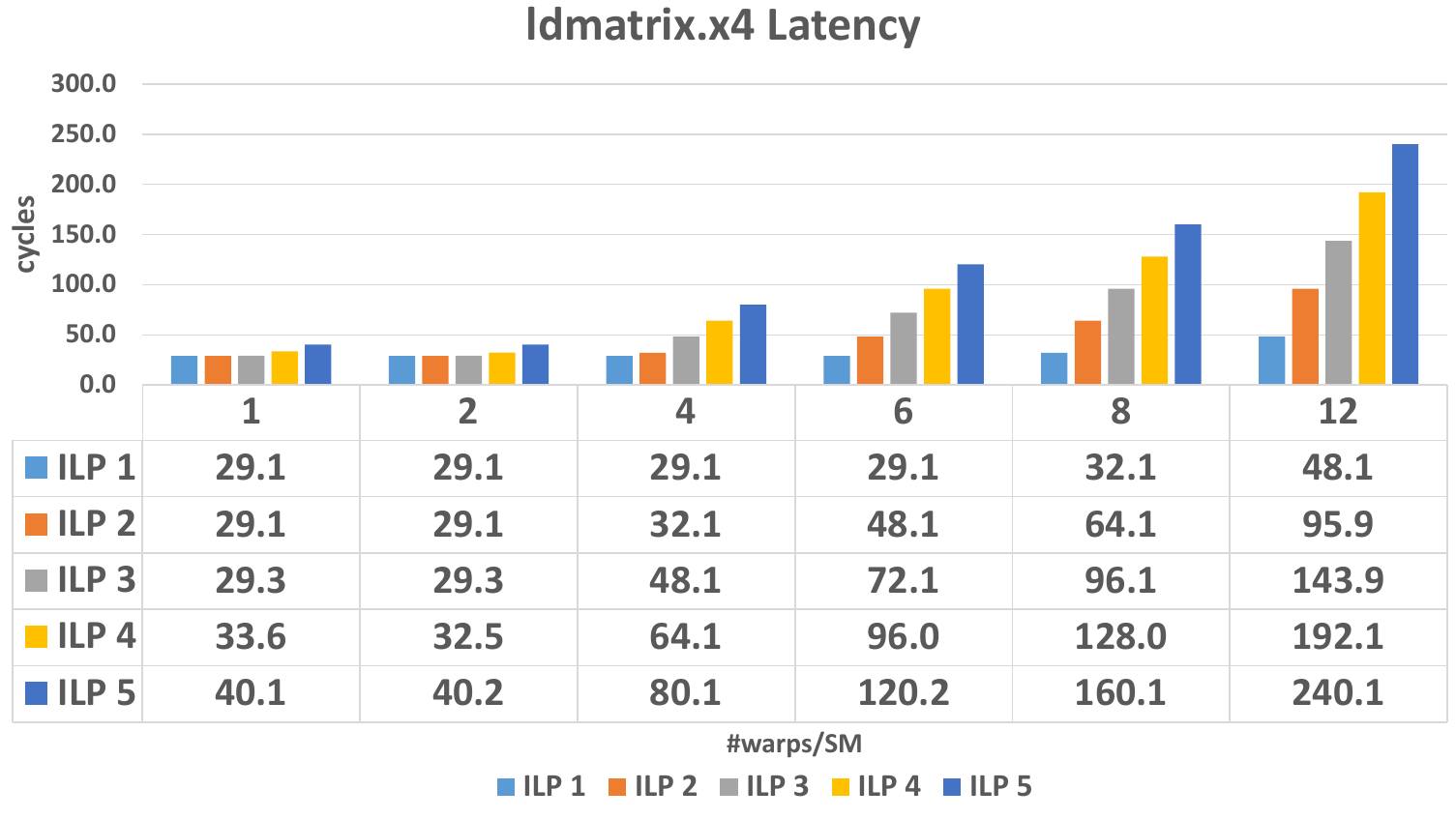}
     \end{subfigure}
\caption{Throughput and latency of ldmatrix.x4 instruction under different settings on A100 Tensor Cores.}
\label{fig:ldmatrix.x4}
\end{figure}

Figure~\ref{fig:ldmatrix.x4} presents the results of ldmatrix.x4 on A100 GPU. Note ldmatrix has three options - ldmatrix.x1, ldmatrix.x2 and ldmatrix.x4 as introduced in Fig.~\ref{fig:func_ldmatrix}. We use ldmatrix.x4 as the example to provide in-depth analyses:
\begin{itemize}
    \item[1)] The completion latency of ldmatrix.x4 is 29 cycles. The measured peak throughput of ldmatrix.x4 can achieve the peak theoretical performance offered by hardware (128 bytes/clk/SM).
    
    \item[2)] If there is only one warp, the peak measured throughput is only around 64 and can be achieved with ILP = 4. To reach the maximum throughput offered by hardware, at least two warps should be used. This indicates there could be two data movement units for ldmatrix.x4 instruction between Shared Memory and Register File. 
    
    \item[3)] Allocating 6 warps will not cause the same problem observed in the experiments of Tensor Cores computation instructions mma and mma.sp. This also confirms that the data movement units are not located in the sub-core.
\end{itemize}


\begin{table*}[htp]
\centering
\caption{Performance of three ldmatrix instructions on A100 GPU.}
\label{table:ldmatrix-summary}
\centering
\small
\setlength\tabcolsep{2pt}
\begin{tabular}{|c|c|c|c|c|c|c|c|c|}
\hline
            & Bytes/warp & \begin{tabular}[c]{@{}c@{}}Completion Latency\\ (cycles)\end{tabular} & (\#warp,ILP) & \begin{tabular}[c]{@{}c@{}}Latency\\ (cycles)\end{tabular} & \begin{tabular}[c]{@{}c@{}}Throughput\\ (bytes/clk/SM)\end{tabular} & (\#warp,ILP) & \begin{tabular}[c]{@{}c@{}}Latency\\ (cycles)\end{tabular} & \begin{tabular}[c]{@{}c@{}}Throughput\\ (bytes/clk/SM)\end{tabular} \\ \hline
ldmatrix.x1 & 128        & 23.1                                                                  & 4,5          & 26.8                                                       & 95.4                                                              & 8,4          & 32.1                                                       & 127.7                                                             \\ \hline
ldmatrix.x2 & 256        & 25.1                                                                  & 4,4          & 32.1                                                       & 127.8                                                             & 8,2          & 32.1                                                       & 127.7                                                             \\ \hline
ldmatrix.x4 & 512        & 29.3                                                                  & 4,2          & 32.2                                                       & 127.3                                                             & 8,1          & 32.6                                                       & 125.9                                                             \\ \hline
\end{tabular}
\end{table*}

Table~\ref{table:ldmatrix-summary} lists the completion latency and performance of two convergence points of three ldmatrix instructions. If choosing ldmatrix.x1 as the data movement instruction, then at least 8 warps should be launched, while for ldmatrix.x2 and ldmatrix.x4, 4 warps are sufficient.

\begin{table}[htp]
\centering
\caption{Latency of ld.shared instructions under different bank conflicts.}
\label{table:ldshared-latency}
\begin{tabular}{|c|c|c|c|c|}
\hline
              & no-conflict & 2-way & 4-way  & 8-way\\ \hline
ld.shared.u32 & 23.0        & 25.0  & 29.0 & 37.0 \\ \hline
ld.shared.u64 & NA          & 25.1  & 29.1 & 37.0 \\ \hline
\end{tabular}
\end{table}

According to the definition of Shared Memory bank conflict, Shared Memory can serve at most 32 threads with a 4 byte/thread access request, which gives 128 byte/clk/warp. Therefore, as shown in Table~\ref{table:ldmatrix-ld.shared}, ldmatrix.x2 and ldmatrix.x4 should intrinsically cause 2-way and 4-way bank conflicts respectively (i.e. two/four memory transactions are needed). However, the NVIDIA official profiler~\cite{NVIDIANsight} does not detect Shared Memory conflicts for both ldmatrix.x2 and ldmatrix.x4 instructions. We further profile the effect of bank conflicts for ld.shared instruction as listed in Table~\ref{table:ldshared-latency}. By comparing results to the completion latency of ldmatrix in Table~\ref{table:ldmatrix-summary}, we have the following observations:

\begin{itemize}
    \item[1)] The completion latency of ldmatrix.x1, ldmatrix.x2, and ldmatrix.x4 are close to the latency of ld.shared.u32 with bank conflict-free, two-way, and four-way bank conflicts respectively. This observation indicates that ldmatrix.x2 and ldmatrix.x4 intrinsically causes bank conflicts and requires more Shared Memory transactions to sequentially serve the memory access. 
    \item[2)] The bank conflict penalty for modern GPUs is around 2 cycles/way, which matches the previous study on older GPUs like Maxwell Architecture~\cite{DissectMemory}.
\end{itemize}

In summary, ldmatrix instructions perform the function of loading data from the Shared Memory to the Register File for Tensor Cores. ldmatrix and ld.shared instructions have similar intrinsic behaviors and performance. The major difference is the layout of destination registers which store the fetched data from Shared Memory. ldmatrix is designed for Tensor Cores computations while ld.shared is preferable for CUDA cores programming.

\section{Numeric behaviors}\label{sec:numeric}
Mixed-precision FMA and low-precision float point data types are becoming increasingly important for Deep Learning applications and some HPC applications. The third generation Ampere Tensor Cores enable tensor computations with three low-precision float point data types: TF32 (19bit), BF16 (16bit), and FP16 (16bit). Table~\ref{table:datatypes} shows different precision formats. Compared to IEEE standard FP32, TF32 and BF16 have the same 8bit exponent but fewer mantissa bits, so they have the same range as FP32 but lower accuracy (fewer mantissa bits). Note that although TF32 only has 19 bits (1+8+10), it stores in the 32-bit register in GPU, which means using TF32 to replace FP32 will not lead to a reduced memory footprint.

\textcolor{Highlight}{Note that we exclude the discussions and experimental results of Integer data type of Tensor Cores, since our experiments show that Integer computations on Tensor Core give 0 errors compared to CPU implementation (i.e. The results of GPU and CPU baseline are the same ) as long as the initialization values are within the data type range (e.g. for INT8, the input data should be initialized within [-128,127]). If the initialization data is out of range, then type casting has to be performed by either a user-defined casting function or C++ default type casting. As long as the casting is the same for the GPU (i.e. Tensor Cores) and CPU baseline, the numeric results are still the same.}

\begin{table}[htp]
\caption{Different precision formats and storage in GPUs. Note that TF32 has 19bits (1+8+10), but it is stored in the 32-bit registers. Two FP8 types will be supported by the forthcoming Hopper Tensor Cores.}
\label{table:datatypes}
\centering
\begin{tabular}{|c|c|c|c|c|}
\hline
data types & sign & exponent & mantissa & register \\ \hline
FP32       & 1    & 8        & 23       & 32b            \\ \hline
TF32       & 1    & 8        & 10       & 32b            \\ \hline
FP16/half   & 1    & 5        & 10       & 16b            \\ \hline
BF16       & 1    & 8        & 7        & 16b          \\ \hline
FP8-E4M3       & 1    & 4        & 3        & 8b          \\ 
\hline
FP8-E5M2       & 1    & 5        & 2        & 8b          \\ 
\hline
\end{tabular}
\end{table}

Although low-precision computation offers significant performance gains as discussed in Section~\ref{sect:DesenTC} and Section~\ref{sect:SparseTC}. Low precision will cause numeric errors compared to FP32 precision. When using low-precision computation to accelerate applications, it is important to understand the numeric behaviors of Tensor Cores. In this Section, we present a set of numeric benchmarks to reveal the properties of low-precision float points operations of Tensor Cores.





\begin{figure}[htp]
     \centering
     \begin{subfigure}[*]{0.23\textwidth}
         \centering
         \includegraphics[width=\textwidth]{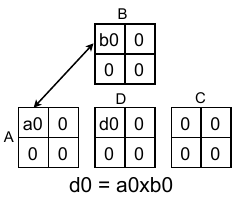}
    \caption{Multiplication}
    \label{fig:mul}
     \end{subfigure}
    \begin{subfigure}[*]{0.23\textwidth}
         \centering
         \includegraphics[width=\textwidth]{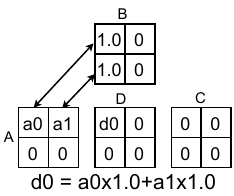}
    \caption{Addition of inner product.}
    \label{fig:addInnerProduct}
     \end{subfigure}
    \begin{subfigure}[*]{0.22\textwidth}
         \centering
         \includegraphics[width=\textwidth]{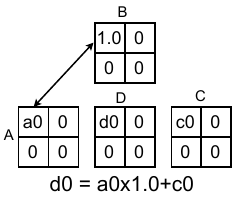}
    \caption{Accumulation.}
    \label{fig:addAccum}
     \end{subfigure}
\caption{Profiling of three operations.}
\end{figure}

\subsection{Element-wise numeric profiling}\label{subsec:element-wise-profiling}

Tensor Cores instruction performs D = A$\times$B + C, which consists of three types of operation: 1) multiplication, 2) addition of inner product, and 3) accumulation. To understand the numeric behavior at the lowest possible level, we use three experiments to profile the three operations separately. 




Figure~\ref{fig:mul},~\ref{fig:addInnerProduct} and~\ref{fig:addAccum} illustrate the ideas of profiling multiplication, addition of inner product, and accumulation, respectively. When profiling multiplication numeric precision, we randomly generate the first element in the first row of matrix A and the first element in the first column of matrix B and set other values as 0.0. Then the matrix FMA D = A$\times$B+C becomes d0 = a0$\times$b0 as shown in Fig.~\ref{fig:mul}. By comparing the result to the FP32 result on CPU as the baseline, we can get the numeric errors of the low-precision floating-point data types used in Tensor Cores. Following the same ideas, we can further profile the addition operation of inner product and addition operation of accumulation. Note that we use a random generator of the normal distribution with $\mu = 0.0$ and $\sigma = 1.0$, and use the same random seed for all experiments so the sequences of random values are the same. This setting helps to compare the precision between different numeric types since they will get the same random initialization values. 

\subsubsection{BF16}


For BF16 Tensor Core operations, the data type for A/B and C/D on Tensor Cores are BF16 and FP32 respectively. We have two initialization strategies, the first is to initialize the data with FP32 precision and convert it to BF16 when copying to GPU. However, this will introduce conversion precision loss. The second approach is to initialize the data with FP16 precision to eliminate the conversion precision loss. After eliminating the conversion loss, we can discover the computation precision used in Tensor Cores by comparing it to the result of FP32 computation precision on CPU\cite{EGEMM-TC}.

\begin{table}[htp]
\caption{The numeric profiling results of BF16 on Tensor Cores w.r.t FP32 on CPU.}
\label{table:BF16-numeric-profiling}
\centering
\begin{tabular}{|c|c|c|}
\hline
                         & init\_BF16 & init\_FP32 \\ \hline
multiplication           & 0.0        &1.29E-03   \\ \hline
add - Inner product & 0.0        & 1.72E-03   \\ \hline
accumulation  & 1.89E-08   & 1.13E-03   \\ \hline
\end{tabular}
\end{table}

Table~\ref{table:BF16-numeric-profiling} shows the average errors under different initialization types. When initializing the data with FP32, all operations have numeric errors. By contrast, when choosing BF16 as the initialization type, multiplication and inner product addition will not cause any numeric errors. This observation indicates that the computation of A$\times$B inside the Tensor Cores is using high-precision~\cite{fasi2021numerical} otherwise it can not give the same result as FP32 computation on the CPU. On the other hand, accumulation error indicates that the accumulation of [A$\times$B] + C is performed with relative low precision.

\subsubsection{FP16}


Half/FP16 Tensor Cores operation offers two data type choices - FP32 and FP16 for accumulation matrix C and result matrix D. 
Instead of just comparing to the FP32 result of CPU, we also compare to the CPU FP16 result converted from CPU FP32 result when using FP16 as the data type for matrix C and matrix D.

\begin{table}[htp]
\caption{The numeric profiling results of FP16 on Tensor Cores w.r.t FP32 on CPU. Matrix C and D are in FP32 data type.}
\label{table:FP16-FP32-numeric-profiling}
\centering
\begin{tabular}{|c|c|c|}
\hline
FP32 as C/D               & init\_FP16 & init\_FP32 \\ \hline
multiplication            & 0.00       & 1.59E-04   \\ \hline
add - Inner Product & 0.00       & 2.18E-04   \\ \hline
accumulation   & 0.00   & 1.36E-04   \\ \hline
\end{tabular}
\end{table}

\begin{table}[htp]
\caption{The numeric profiling results of FP16 on Tensor Cores w.r.t FP32 on CPU. Matrix C and D are in FP16 data type.}
\label{table:FP16-FP16-numeric-profiling}
\centering
\addtolength{\tabcolsep}{-2pt}
\begin{tabular}{|c|cc|cc|}
\hline
                          & \multicolumn{2}{c|}{CPU\_FP32}               & \multicolumn{2}{c|}{CPU\_FP32cvtFP16}        \\ \hline
FP16 as C/D               & \multicolumn{1}{c|}{init\_FP16} & init\_FP32 & \multicolumn{1}{c|}{init\_FP16} & init\_FP32 \\ \hline
multiplication            & \multicolumn{1}{c|}{1.22E-04}   & 1.94E-04   & \multicolumn{1}{c|}{0.00}       & 1.67E-04   \\ \hline
add - inner product & \multicolumn{1}{c|}{1.81E-04}   & 2.99E-04   & \multicolumn{1}{c|}{0.00}       & 2.21E-04   \\ \hline
accumulation   & \multicolumn{1}{c|}{1.81E-04}   & 2.99E-04   & \multicolumn{1}{c|}{0.00}       & 2.21E-04   \\ \hline
\end{tabular}
\end{table}

Table~\ref{table:FP16-FP32-numeric-profiling} shows the profiling results of using FP32 as the data type of matrices C and D. When initializing with FP16, there are no numeric errors for all three operations. When initializing with FP32, numeric errors occur due to type conversion like the observations in BF16 profiling. However, the error level(E-04) of FP16 is lower than the error level of BF16 (E-03) since FP16 has more mantissa bits and is more accurate as long as the values are within the FP16 range.

Table~\ref{table:FP16-FP16-numeric-profiling} shows the profiling results of using FP16 as the data type of matrix C and D. Since the result matrix D is in FP16, there are always numeric errors when compared to the CPU FP32 baseline. Interestingly, if we compare the CPU FP16 results, the errors are zero when using FP16 as the initialization data type. This observation suggests that even though the matrix D is in FP16, the hardware conducts the computation in high-precision and only converts the final result to low-precision FP16 in the end. 


\subsubsection{TF32}

\begin{table}[htp]
\caption{The numeric profiling results of TF32 on Tensor Cores w.r.t FP32 on CPU.}
\label{table:TF32-numeric-profiling}
\centering
\begin{tabular}{|c|c|c|}
\hline
              & init\_TF32 & init\_FP32 \\ \hline
multiplication    & 0.0   & 1.59E-04   \\ \hline
add - Inner Product & 0.0   & 2.17E-04   \\ \hline
accumulation  & 0.0   & 1.36E-04   \\ \hline
\end{tabular}
\end{table}

The profiling code for TF32 Tensor Cores is similar to the one for the BF16 data type. Table~\ref{table:TF32-numeric-profiling} presents the results of TF32 Tensor Cores operations. Note that if we compare FP16 results in Table~\ref{table:FP16-FP32-numeric-profiling}, we can find that TF32 and FP16 give the same level of error because they have the same number of mantissa bits (10 bits).

\subsection{Chain Matrix Multiplication}

We use a simple application - chain matrix multiplication, to evaluate the effect of low-precision FMA. Chain matrix multiplication is a simplified computation pattern of modern deep learning applications which consists of many layers and the results of the previous layer will be the input of the next layer. In this computation pattern, both range (exponent bits) and precision (mantissa bits) are important and will affect the final accuracy. Fewer mantissa bits will result in larger accumulative errors for the final result of the chain. Fewer exponent bits give a smaller valid range and will result in overflow (infinity) earlier. 



The initialization data type can be either FP32 or low-precision type (BF16/FP16). For each node of the chain, we first compute D = A$\times$B. The result matrix D will be assigned to matrix A for the next computation, and matrix B will be assigned new random values. Unlike the element-wise profiling in Section~\ref{subsec:element-wise-profiling} where we only evaluate the errors of one element, we evaluate the numeric errors of the whole matrix D with shape m$\times$n. Based on our study in Section~\ref{sect:DesenTC}, we choose mma instructions with shape m16b8k8 since this common shape is supported by BF16, FP16, and TF32. After completion the chain computations, we evaluate the l2 relative errors between CPU results $D^{FP32}$ and Tensor Cores low-precision results $D^{l}$ with equation~\ref{eq:relativeErr}.

\begin{equation}\label{eq:relativeErr}
    RelativeErr = \frac{ \sqrt {\sum_{i=1}^{m}\sum_{j=1}^{n} |D^{l}_{ij} - D^{FP32}_{ij}|^2 }}{\sqrt{ \sum_{i=1}^{m}\sum_{j=1}^{n}|D^{l}_{ij}|^2}}
\end{equation}

\begin{figure}[htp]
     \centering
     \begin{subfigure}[*]{0.485\textwidth}
         \centering
         \includegraphics[width=\textwidth]{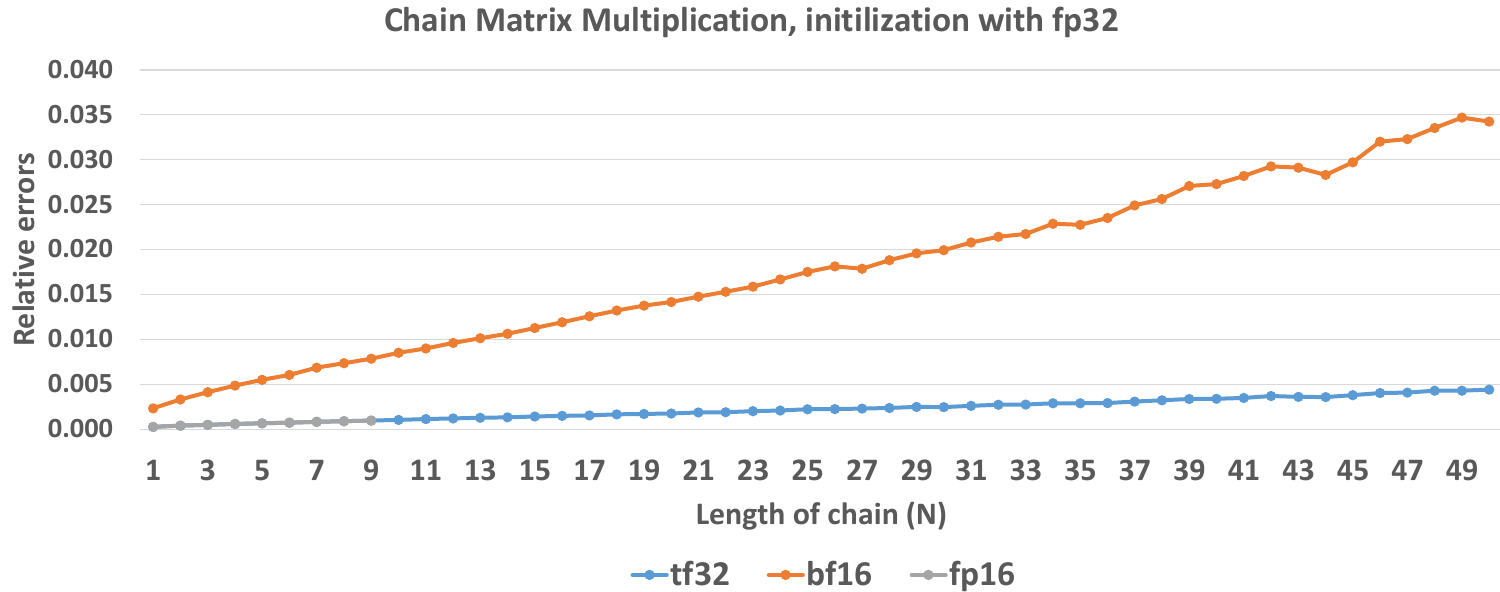}
     \end{subfigure}
    \begin{subfigure}[*]{0.485\textwidth}
         \centering
         \includegraphics[width=\textwidth]{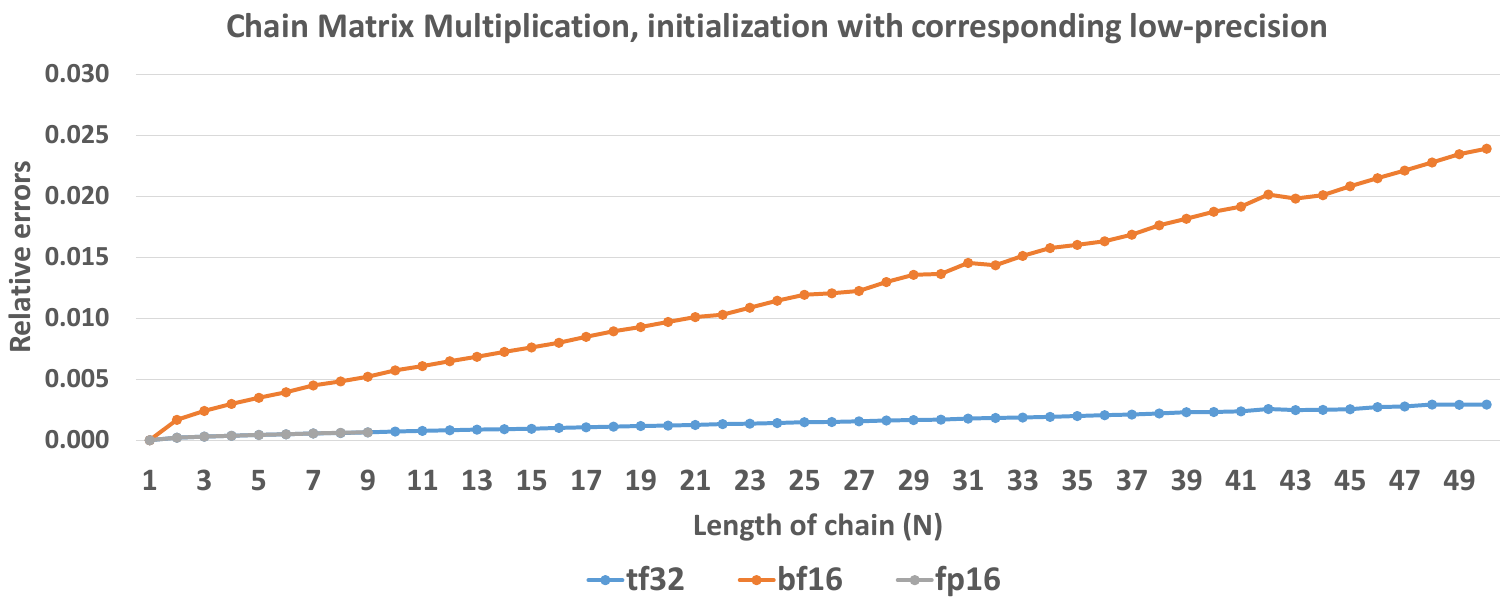}
     \end{subfigure}
\caption{Numeric profiling of chain matrix multiplication with different data types -TF32, BF16, and FP16. All values are randomly generated by normal distribution with $\mu = 0$ and $\sigma = 1$. The errors are taken on average with 1000 measurements. Note that the line of FP16 stops at N = 10 due to overflow (infinity).}
\label{fig:chain-result}
\end{figure}

Figure~\ref{fig:chain-result} shows the results of three data types of chain matrix multiplication with different chain lengths (N). We have the following observations:  
\begin{itemize}
    \item[1)] In general the errors increase with the length of the chain. This increase is more significant for the BF16 data type compared to TF32 and FP16, this is because BF16 has fewer mantissa bits and will suffer more from accumulative errors.
    \item[2)] The results of FP16 give the same error levels as using TF32 because FP16 and TF32 have the same number of mantissa bits which gives the same level of precision, but it will run into overflow (infinity) after N $\geq$ 10 since it has fewer exponent bits which result in a less range of numbers can be represented.
    \item[3)] Using FP32 as an initialization strategy always causes higher errors since it introduces type conversion precision loss. When using a low-precision type for initialization, the error is almost zero when chain length is one because the internal computations are conducted in high-precision as discovered in Section~\ref{subsec:element-wise-profiling} and there is no conversion precision loss. 
\end{itemize}

In summary, FP16 offers the same level of precision as TF32 as long as the numbers are within the valid range that can be represented by FP16. BF16 has the same valid range as TF32 but it suffers more seriously from accumulative precision loss. The users should consider both numeric behaviors and the performance benefits studied in Section~\ref{sect:DesenTC} and ~\ref{sect:SparseTC} when choosing the target data type for accelerating their applications on Tensor Cores.

\section{Conclusion}\label{sec:conclusion}


We dissect NVIDIA Tensor Cores, especially the newest Ampere Tensor Cores through a set of microbenchmarks and numeric profiling experiments. We summarize our main findings as follows:

\begin{itemize}
    \item When programming Ampere Tensor Cores. Using the new ldmatrix + mma instructions is favorable compared to the legacy wmma.load + wmma.mma instructions. \textcolor{Highlight}{The ablation experiments in the Appendix \ref{appendix:ablation} demonstrate the advantages of using the new instructions. Specifically, by levering the flexibility of the new ldmatrix instruction, it can reduce the number of corresponding GPU clock cycles by more than 60\%} .   

    \item Sparse acceleration introduced in Ampere Tensor Cores can only be programmed through mma APIs. Sparse operation doubles the throughput by accepting larger input matrices than dense ones while using the same number of execution cycles. 
    
    \item It is well known that four warps should be allocated on a GPU SM since there are four warp schedulers~\cite{AmpereWhite,raihan2019modeling}. \textcolor{Highlight}{ Our experiments show that for some instructions (e.g. Fig.~\ref{fig:A100-m16n8k8}), peak performance can only be achieved when there are at least eight warps} .
    
    \item For each data type, there are two mma.sp instructions with different k sizes. In general, the instructions with larger k can achieve the expected performance on both data-center A100 and Gaming RTX3070Ti Tensor Cores. However, the instructions with smaller k give an undesired performance on A100 Tensor Cores (i.e. can not achieve peak throughput).
    
    \item The performance of different GPUs with the same Architecture generation may not be consistent. For instance, RTX3070Ti Tensor Cores favor FP16 as an accumulation data type from a performance perspective, but there is no difference no matter whether using FP16 or FP32 on A100 Tensor Cores. 
  
    \item \textcolor{Highlight}{The performance of BF16 and FP16 instructions on Tensor Cores are the same, which confirms the observations in previous work~\cite{fasi2021numerical}. On the other hand, our experiments intuitively show that FP16 suffers from a smaller range and BF16 suffers from higher numeric errors. For machine learning applications that have a relatively high tolerance for numeric precision loss, BF16 is favorable since it has the same range as F32 and therefore the same overflow behavior. }
\end{itemize}

In conclusion, our work provides comprehensive and up-to-date information on the Tensor Cores, especially the instruction-level evaluations.  To the best of our knowledge, this paper is the first systematic study on recent Tensor Cores generations (Turing and Ampere) using the new programming interface. We believe our contributions can help users to implement their own Tensor Cores applications and facilitate accurate Tensor Cores modeling.

\ifCLASSOPTIONcompsoc
  \section*{Acknowledgments}
\else
  \section*{Acknowledgment}
\fi

This work is funded by the Dutch Research Council (NWO) Perspectief Program ZERO-ARM P3. This work is also supported by the U.S. Department of Energy, Office of Science, Office of Advanced Scientific Computing Research, ComPort: Rigorous Testing Methods to Safeguard Software Porting, under Award Number 78284. The Pacific Northwest National Laboratory is operated by Battelle for the U.S. Department of Energy under Contract DE-AC05-76RL01830.






%




\bibliographystyle{IEEEtranS}
\bibliography{bare_jrnl_compsoc}

\color{Highlight}
\appendices
\section{Ablation experiments}\label{appendix:ablation}


In this section, we use three matrix multiplication CUDA kernels to demonstrate the advantages of using the new instructions studied in this paper. 
\subsection{Benefits of shared memory usage}\label{appendix:sharedmemUsage}
In section \ref{sect:DataMovement}, we excluded the experiments for data movement from global memory to shared memory and assume the data are ready in shared memory for ldmatrix instructions which can only fetch data from shared memory to registers. We explained in section \ref{sec:Background} with two reasons:
\begin{itemize}
    \item[1] Using shared memory as the buffer to reduce global memory traffic and increase data reuse.
    \item[2] The novel asynchronous memory copy introduced in Ampere Architecture facilitates the software data pipeline by using shared memory as the staging storage to overlap the data movement with computation.
\end{itemize}
The first point - using shared memory to increase data reuse has been widely used and been a fundamental optimization technique. A detailed study of this technique can be found in textbook~\cite{kirk2016programming} so we exclude further discussions here and emphasize the second point - new asynchronous memory copy.

Asynchronous memory copy acceleration was introduced in Ampere Architecture~\cite{AmpereWhite}. It allows asynchronous data movement from off-chip global memory to on-chip shared memory. Compared to the old synchronous copy fashion, asynchronous copy can be leveraged to hide the data copy latency by overlapping it with computation. To demonstrate the advantages of  asynchronous copy, we implemented two matrix multiplication CUDA kernels.
\begin{itemize}
    \item mma\_baseline.cu
    \begin{itemize}
        \item[a] Copy the data tile (among inner dimension K) from global memory to shared memory.
        \item[b] Synchronization. Wait until all copy ready to avoid data hazards.
        \item[c] Copy data from shared memory to register with ldmatrix.
        \item[d] Use Tensor Core (mma) to do the computations.
        \item[e] Repeat a-d until the end of the K dimension.
    \end{itemize}
\end{itemize}

\begin{itemize}
    \item mma\_pipeline.cu
    \begin{itemize}
        \item[a] Copy the first data tile from global memory to shared memory with the asynchronous copy.
        \item[b] Copy the next data tile from global memory to shared memory with the asynchronous copy.
        \item[c] Wait until the first tile copy is complete.
        \item[d] Copy data from shared memory to register with ldmatrix.
        \item[e] Use Tensor Core (mma) to do the computation.
        \item[f] Repeat a-e until the end of the K dimension.
    \end{itemize}
\end{itemize}

We profile the number of cycles (using clock64() CUDA API) of the above two implementations with large matrix multiplication (2048 x 2048 with bf16 data type).

\begin{table}[htp]
\caption{\textcolor{Highlight}{Compare baseline with asynchronous copy pipeline. Tested on an Ampere A100 GPU.}}
\label{table:Asynchronous pipeline}
\centering
\begin{tabular}{|c|c|}
\hline
Implementation   & Number of GPU cycles \\ \hline
mma\_baseline.cu & 913363           \\ \hline
mma\_pipeline.cu & 451560           \\ \hline
\end{tabular}
\end{table}

The results demonstrate the advantages of the asynchronous copy pipeline. Therefore we can conclude that shared memory should be used for the best possible performance.

\subsection{Flexibility of ldmatrix instruction}\label{appendix:flexibilityOfLdmatrix}

As introduced in Figure \ref{fig:ld_gra} of section \ref{sect:DataMovement}. ldmatrix instruction has more flexibility than the legacy wmma.load instruction, which helps reduce bank conflicts by using some special shared data layout. For instance, CUTLASS introduced permuted shared memory layout~\cite{CUTLASS-Permuted} to reduce bank conflicts. We followed the ideas and re-implemented our third CUDA kernel mma\_permuted.cu by adjusting the mma\_baseline.cu implementation. Table \ref{table:Permuted shared memory} shows the profiling results.

\begin{table}[htp]
\caption{\textcolor{Highlight}{Compare baseline with permuted shared memory layout. Tested on an Ampere A100 GPU.}}
\label{table:Permuted shared memory}
\centering
\begin{tabular}{|c|c|}
\hline
Implementation   & Number of GPU cycles \\ \hline
mma\_baseline.cu & 913363           \\ \hline
mma\_permuted.cu & 303227           \\ \hline
\end{tabular}
\end{table}

The results show the advantages of leveraging the flexibility of ldmatrix to reduce bank conflicts, which can therefore reduce the number of cycles significantly.

Note that ldmatrix should always be used with mma instructions since ldmatrix will store the fetched data in agreement with the requirements of mma instructions as specified by vendor's documentations~\cite{NVPTX}. In another word, the correct execution chain should be either ldmatrix + mma or the legacy wmma.load + wmma.mma.





\end{document}